\def\complexNumbers{\mathbb{C}}
\def\realNumbers{\mathbb{R}}

\def\constante{{\rm e}}

\def\expectationOperator[#1][#2]{{\mathbb{E}_{#2}}\left[#1\right]}
\def\indicatorFunction[#1]{\mathbb{I}\left[{#1}\right]}
\def\probability[#1]{\mathbb{P}\left[{#1}\right]}

\def\numberOfEdgeDevices{K}
\def\timeDomainOFDM[#1]{x(#1)}
\def\timeVar{t_{\rm time}}

\def\indexSubcarrier{l}
\def\numberOfActiveSubcarriers{M}

\def\dataSymbols[#1]{d_{#1}}

\def\symbolDuration{T_{\rm s}}
\def\Ptransmit{P_{\rm tx}}
\def\receivedSymbolAtSubcarrier[#1]{r_{#1}^{(\indexCommunicationRound)}}
\def\transmittedSymbolAtSubcarrier[#1]{t_{#1}^{(\indexCommunicationRound)}}
\def\randomSymbolAtSubcarrier[#1]{s_{#1}^{(\indexCommunicationRound)}}
\def\channelAtSubcarrier[#1]{h_{#1}^{(\indexCommunicationRound)}}
\def\noiseAtSubcarrier[#1]{n_{#1}^{(\indexCommunicationRound)}}
\def\numberOfOFDMSymbols{S}
\def\indexOFDMSymbol{m}
\def\asymbolFromED[#1]{d_{#1}}

\def\exponentialIntegral[#1]{{\rm Ei}(#1)}
\def\tciFactor[#1]{p_{#1}}
\def\mappingFunction{{f}}
\def\encoder[#1]{\psi_#1}
\def\symbolEnergy{E_{\rm s}}

\def\voteInTime[#1]{m^{#1}}
\def\voteInFrequency[#1]{l^{#1}}

\def\numberOFEDsForOptionOne{K^{+}_{\indexGradient}}
\def\numberOFEDsForOptionSecond{K^{-}_{\indexGradient}}
\def\noiseVariance{\sigma_{\rm n}^2}

\def\coefficientOne{a}

\def\correctDecision[#1]{p_{#1}}
\def\incorrectDecision[#1]{q_{#1}}
\def\aparameterForBer[#1]{\epsilon_{#1}}
\def\numberOfEDsWithCorrectChoice{Z}
\def\probabilityIncorrect[#1]{P^{\rm err}_{#1}}
\def\meanOptionOne{\mu^{+}_{\indexGradient}}
\def\meanOptionTwo{\mu^{-}_{\indexGradient}}
\def\effectiveSNR{\xi}

\def\identiyVector[#1]{\textbf{\textrm{I}}_{#1}}
\def\zeroVector[#1]{\textbf{\textrm{0}}_{#1}}

\def\dataset[#1]{\mathcal{D}_{#1}}

\def\datasetBatch[#1]{\mathcal{\tilde{D}}_{#1}}
\def\batchSize{n_{\rm b}}
\def\brachSizeRelativeToRounds{\gamma}
\def\completeData{\mathcal{D}}
\def\numberOfModelParameters{q}
\def\sampleData[#1]{{\textrm{\textbf{x}}}_{#1}}
\def\sampleLabel[#1]{{y}_{#1}}
\def\learningRate{\eta}

\def\deltaVectorAtIteration[#1][#2]{{\Delta}^{(#1)}_{#2}}
\def\deltaVectorAtIterationEle[#1]{{\bm \Delta}^{#1}}

\def\indexED{k}
\def\indexGradient{i}
\def\indexSampleData{{\ell}}
\def\indexCommunicationRound{n}

\def\modelParametersAtIteration[#1]{\textbf{w}^{(#1)}}
\def\modelParametersAtIterationEle[#1][#2]{w^{(#1)}_{#2}}

\def\modelParameters{\textbf{w}}
\def\modelParametersEle[#1]{{w}_{#1}}
\def\modelParametersOptimal{{\textbf{w}^{*}}}

\def\localGradientSign[#1][#2]{\bar{\textbf{g}}_{#1}^{(#2)}}
\def\localGradient[#1][#2]{\tilde{\textbf{g}}_{#1}^{(#2)}}
\def\localGradientNoIndex[#1]{\tilde{\textbf{g}}_{#1}}

\def\localGradientSignElement[#1][#2]{{\bar{g}}_{#1}^{(#2)}}
\def\localGradientElement[#1][#2]{{\tilde{g}}_{#1}^{(#2)}}
\def\localGradientNoIndexElement[#1]{{\tilde{g}}_{#1}}

\def\lossFunctionSample[#1]{f(#1)}
\def\lossFunctionLocal[#1][#2]{F_{#1}(#2)}
\def\lossFunctionGlobal[#1]{F(#1)}
\def\lossFunctionGlobalMinimum{F^*}

\def\majorityVoteEle[#1][#2]{{v}^{(#1)}_{#2}}
\def\majorityVote[#1]{\textbf{v}^{(#1)}}

\def\globalGradient[#1]{{\textbf{{g}}}^{(#1)}}
\def\globalGradientElement[#1][#2]{{{g}}^{(#1)}_{#2}}
\def\globalGradientNoIndex{{\textbf{{g}}}}
\def\globalGradientElementNoIndex[#1]{{g_{#1}}}

\def\communicationRounds{N}

\def\metricForFirst[#1]{e_{#1}^{+}}
\def\metricForSecond[#1]{e_{#1}^{-}}

\def\nonnegativeConstants{\textbf{L}}
\def\nonnegativeConstantsEle[#1]{L_{#1}}
\def\varianceBound{{\bm \sigma}}
\def\varianceBoundEle[#1]{\sigma_{#1}}

\def\symbolVector[#1]{\textbf{\textrm{d}}_{#1}}
\def\symbolVectorEstimate[#1]{\tilde{\textbf{\textrm{{d}}}}_{#1}}
\def\receivedVector[#1]{\textbf{\textrm{r}}_{#1}}
\def\noiseVector[#1]{\textbf{\textrm{n}}_{#1}}

\def\transmittedVector[#1]{\textbf{\textrm{t}}_{#1}}
\def\idftMatrix[#1]{\textbf{\textrm{F}}_{#1}^{\rm H}}
\def\dftMatrix[#1]{\textbf{\textrm{F}}_{#1}}
\def\transformPrecoder[#1]{\textbf{\textrm{D}}_{#1}}
\def\transformDecoder[#1]{\textbf{\textrm{D}}_{#1}^{\rm H}}

\def\channelMatrix[#1]{\textbf{\textrm{H}}_{#1}}

\def\syncError{T_{\rm sync}}

\def\referenceDistance{R_{\rm ref}}
\def\minimumDistance{R_{\rm min}}
\def\cellRadius{R_{\rm max}}
\def\distanceED[#1]{r_{#1}}
\def\powerED[#1]{P_{#1}}
\def\pathlossExponent{\alpha}
\def\powerControl{\beta}
\def\effectivePathLossExponent{\alpha_{\rm eff}}
\def\largeScaleImpactOnLearning{\lambda}
\def\arandomvar{y}
\def\indexArea{u}
\def\Nerror{N_{\text{err}}}

\documentclass[conference]{IEEEtran}
\usepackage{multicol}
\usepackage{lipsum}
\usepackage{mathtools}
\usepackage{amsthm}
\usepackage{acronym}
\usepackage{stfloats}
\usepackage{amsfonts}
\usepackage{acronym}
\usepackage{cite}
\usepackage{multirow}
\usepackage{bm}
\usepackage{amsmath,amssymb}
\usepackage{graphicx}
\usepackage[table,xcdraw]{xcolor}
\usepackage[caption=false,font=footnotesize]{subfig}
\usepackage[geometry]{ifsym}
\usepackage{array}
\usepackage{flushend}
\usepackage[utf8]{inputenc}
\usepackage[T1]{fontenc}

\usepackage[cmintegrals]{newtxmath}

\newcommand\mydots{\hbox to 1em{.\hss.\hss.}}
\let\norm\undefined 
\DeclarePairedDelimiter\norm{\lVert}{\rVert}

\usepackage{tikz}
\usepackage{lipsum}
\usetikzlibrary{calc,shadings,patterns}
\makeatletter
\tikzset{%
  remember picture with id/.style={%
    remember picture,
    overlay,
    save picture id=#1,
  },
  save picture id/.code={%
    \edef\pgf@temp{#1}%
    \immediate\write\pgfutil@auxout{%
      \noexpand\savepointas{\pgf@temp}{\pgfpictureid}}%
  },
  if picture id/.code args={#1#2#3}{%
    \@ifundefined{save@pt@#1}{%
      \pgfkeysalso{#3}%
    }{
      \pgfkeysalso{#2}%
    }
  }
}

\def\savepointas#1#2{%
  \expandafter\gdef\csname save@pt@#1\endcsname{#2}%
}

\def\tmk@labeldef#1,#2\@nil{%
  \def\tmk@label{#1}%
  \def\tmk@def{#2}%
}

\tikzdeclarecoordinatesystem{pic}{%
  \pgfutil@in@,{#1}%
  \ifpgfutil@in@%
    \tmk@labeldef#1\@nil
  \else
    \tmk@labeldef#1,(0pt,0pt)\@nil
  \fi
  \@ifundefined{save@pt@\tmk@label}{%
    \tikz@scan@one@point\pgfutil@firstofone\tmk@def
  }{%
  \pgfsys@getposition{\csname save@pt@\tmk@label\endcsname}\save@orig@pic%
  \pgfsys@getposition{\pgfpictureid}\save@this@pic%
  \pgf@process{\pgfpointorigin\save@this@pic}%
  \pgf@xa=\pgf@x
  \pgf@ya=\pgf@y
  \pgf@process{\pgfpointorigin\save@orig@pic}%
  \advance\pgf@x by -\pgf@xa
  \advance\pgf@y by -\pgf@ya
  }%
}

\makeatother

\newcounter{hatchNumber}
\usepackage{cuted}
\makeatletter
\newif\ifAC@uppercase@first%
\def\Aclp#1{\AC@uppercase@firsttrue\aclp{#1}\AC@uppercase@firstfalse}%
\def\AC@aclp#1{%
	\ifcsname fn@#1@PL\endcsname%
	\ifAC@uppercase@first%
	\expandafter\expandafter\expandafter\MakeUppercase\csname fn@#1@PL\endcsname%
	\else%
	\csname fn@#1@PL\endcsname%
	\fi%
	\else%
	\AC@acl{#1}s%
	\fi%
}%
\def\Acp#1{\AC@uppercase@firsttrue\acp{#1}\AC@uppercase@firstfalse}%
\def\AC@acp#1{%
	\ifcsname fn@#1@PL\endcsname%
	\ifAC@uppercase@first%
	\expandafter\expandafter\expandafter\MakeUppercase\csname fn@#1@PL\endcsname%
	\else%
	\csname fn@#1@PL\endcsname%
	\fi%
	\else%
	\AC@ac{#1}s%
	\fi%
}%
\def\Acfp#1{\AC@uppercase@firsttrue\acfp{#1}\AC@uppercase@firstfalse}%
\def\AC@acfp#1{%
	\ifcsname fn@#1@PL\endcsname%
	\ifAC@uppercase@first%
	\expandafter\expandafter\expandafter\MakeUppercase\csname fn@#1@PL\endcsname%
	\else%
	\csname fn@#1@PL\endcsname%
	\fi%
	\else%
	\AC@acf{#1}s%
	\fi%
}%
\def\Acsp#1{\AC@uppercase@firsttrue\acsp{#1}\AC@uppercase@firstfalse}%
\def\AC@acsp#1{%
	\ifcsname fn@#1@PL\endcsname%
	\ifAC@uppercase@first%
	\expandafter\expandafter\expandafter\MakeUppercase\csname fn@#1@PL\endcsname%
	\else%
	\csname fn@#1@PL\endcsname%
	\fi%
	\else%
	\AC@acs{#1}s%
	\fi%
}%
\edef\AC@uppercase@write{\string\ifAC@uppercase@first\string\expandafter\string\MakeUppercase\string\fi\space}%
\def\AC@acrodef#1[#2]#3{%
	\@bsphack%
	\protected@write\@auxout{}{%
		\string\newacro{#1}[#2]{\AC@uppercase@write #3}%
	}\@esphack%
}%
\def\Acl#1{\AC@uppercase@firsttrue\acl{#1}\AC@uppercase@firstfalse}
\def\Acf#1{\AC@uppercase@firsttrue\acf{#1}\AC@uppercase@firstfalse}
\def\Ac#1{\AC@uppercase@firsttrue\ac{#1}\AC@uppercase@firstfalse}
\def\Acs#1{\AC@uppercase@firsttrue\acs{#1}\AC@uppercase@firstfalse}

\newtheorem{theorem}{Theorem}

\newtheorem{lemma}{Lemma}

\newtheorem{assumption}{Assumption}

\DeclareMathOperator{\sign}{sign}
\def\signNormal[#1]{\sign\left(#1\right)}

\acrodef{SNR}{signal-to-noise ratio}
\acrodef{RMSE}{root-mean-square error}
\acrodef{OFDM}{orthogonal frequency division multiplexing}
\acrodef{DFT}{discrete Fourier transform}
\acrodef{PSK}{phase-shift keying}
\acrodef{QAM}{quadrature amplitude modulation}
\acrodef{QPSK}{quadrature phase-shift keying}
\acrodef{PMEPR}{peak-to-mean envelope power ratio}
\acrodef{BER}{bit-error ratio}
\acrodef{SNR}{signal-to-noise ratio}
\acrodef{PSD}{power spectral density}
\acrodef{SE}{spectral efficiency}
\acrodef{CP}{cyclic prefix}
\acrodef{AWGN}{additive white Gaussian noise}
\acrodef{CFR}{channel frequency response}
\acrodef{CIR}{channel impulse response}
\acrodef{MMSE}{minimum mean square error}
\acrodef{LMMSE}{linear minimum mean square error}
\acrodef{BPSK}{binary phase shift keying}
\acrodef{BLER}{block-error rate}
\acrodef{ML}{maximum likelihood}
\acrodef{PHY}{physical layer}
\acrodef{PA}{power amplifier}
\acrodef{IDFT}{inverse DFT}
\acrodef{DoF}{degrees-of-freedom}
\acrodef{IoT}{Internet-of-Things}
\acrodef{FDE}{frequency-domain equalization}
\acrodef{FSK}{frequency-shift keying}
\acrodef{FSK-MV}{\ac{FSK}-based \ac{MV}}
\acrodef{RF}{radio-frequency}
\acrodef{IM}{index modulation}
\acrodef{BS}{base station}
\acrodef{MF}{matched filter}

\acrodef{BAA}{broadband analog aggregation}
\acrodef{OBDA}{one-bit broadband digital aggregation}
\acrodef{FEEL}{federated edge learning}
\acrodef{FL}{federated learning}
\acrodef{ED}{edge device}
\acrodef{ES}{edge server}
\acrodef{UL}{uplink}
\acrodef{DL}{downlink}
\acrodef{OAC}[AirComp]{over-the-air computation}
\acrodef{TCI}{truncated-channel inversion}
\acrodef{MV}{majority vote}
\acrodef{CNN}{convolution neural network}
\acrodef{ReLU}{rectified-linear unit}
\acrodef{CSI}{channel state information}
\acrodef{PAPR}{peak-to-average power ratio}
\acrodef{iid}[IID]{independent and identically distributed}
\acrodef{5G}{Fifth Generation}
\acrodef{4G}{Fourth Generation}
\acrodef{NR}{New Radio}
\acrodef{LTE}{Long Term Evolution}
\acrodef{RACH}{random-access channel}
\acrodef{DNN}{deep nueral network}
\acrodef{SGD}{stochastic gradient descend}
\acrodef{SGD}{stochastic gradient descend}
\acrodef{signSGD}{sign stochastic gradient descend}

\acrodef{5G}{Fifth Generation}
\acrodef{4G}{Fourth Generation}
\acrodef{NR}{New Radio}
\acrodef{LTE}{Long Term Evolution}
\acrodef{PRACH}{physical random access channel}
\acrodef{PUCCH}{physical uplink control channel}
\acrodef{OFDMA}{orthogonal frequency division multiple access}

\def\BibTeX{{\rm B\kern-.05em{\sc i\kern-.025em b}\kern-.08em
    T\kern-.1667em\lower.7ex\hbox{E}\kern-.125emX}}
\begin{document}

\title{Distributed Learning over a Wireless Network with FSK-Based Majority Vote}

\author{
	\IEEEauthorblockN{Alphan \c{S}ahin}
	\IEEEauthorblockA{Electrical  Engineering Department\\
	University of South Carolina\\
	Columbia, SC, USA\\
	Email: asahin@mailbox.sc.edu}	
		\and	
	\IEEEauthorblockN{Bryson Everette}
	\IEEEauthorblockA{Electrical  Engineering Department\\
		University of South Carolina\\
		Columbia, SC, USA\\
		Email: everetb@email.sc.edu}
		\and
	\IEEEauthorblockN{Safi Shams Muhtasimul Hoque}
	\IEEEauthorblockA{Electrical  Engineering Department\\
	University of South Carolina\\
	Columbia, SC, USA\\
	Email: shoque@email.sc.edu}
}
\makeatletter
\def\ps@IEEEtitlepagestyle{%
	\def\@oddfoot{\mycopyrightnotice}%
	\def\@oddhead{\hbox{}\@IEEEheaderstyle\leftmark\hfil\thepage}\relax
	\def\@evenhead{\@IEEEheaderstyle\thepage\hfil\leftmark\hbox{}}\relax
	\def\@evenfoot{}%
}
\def\mycopyrightnotice{%
	\begin{minipage}{\textwidth}
		\centering \scriptsize
		\copyright~2021 IEEE. Personal use of this material is permitted.  Permission from IEEE must be obtained for all other uses, in any current or future media, including reprinting/republishing this material for advertising or promotional purposes, creating new collective works, for resale or redistribution to servers or lists, or reuse of any copyrighted component of this work in other works.
	\end{minipage}
}
\makeatother

\maketitle

\begin{abstract}
In this study, we propose an \ac{OAC} scheme for \ac{FEEL}. 
 The proposed scheme relies on the concept of distributed learning by  \ac{MV} with \ac{signSGD}. As compared to the state-of-the-art solutions, with the proposed method, \acp{ED} transmit the signs of local stochastic gradients by activating one of two orthogonal resources, i.e., \ac{OFDM} subcarriers, and the \acp{MV} at the \ac{ES} are obtained with non-coherent detectors by exploiting the energy accumulations on the subcarriers. Hence, the proposed scheme eliminates the need for \ac{CSI} at the \acp{ED} and \ac{ES}.  By taking path loss, power control, cell size, and the probabilistic nature of the detected \acp{MV} in fading channel  into account, we prove the convergence of the distributed learning for a non-convex function.
Through simulations, we show that the proposed scheme can provide  a high test accuracy in fading channels even when the time-synchronization and the power alignment at the \ac{ES} are not ideal. We also provide insight into distributed learning for  location-dependent data distribution for the \ac{MV}-based schemes.

\end{abstract}
\section{Introduction}

\acresetall
\Ac{FEEL} is an implementation of \ac{FL} in a wireless network to train a model without moving the local data generated at the \acp{ED} to an \ac{ES} \cite{gafni2021federated, chen2021distributed}. With \ac{FEEL}, a large number of model parameters (or gradients) needs to be communicated between many \acp{ED} and the \ac{ES} through wireless channels. However, typical user multiplexing methods such as \ac{OFDMA} can be inefficient to address the spectrum congestion due  to a large number of \acp{ED} \cite{hellstrom2020wireless}.  To address this issue, one of the promising solutions is to perform the calculations needed for \ac{FEEL}, e.g., averaging, with an \ac{OAC} method that harnesses the signal-superposition property of the wireless-multiple access channel \cite{Goldenbaum_2013,Wanchun_2020,Nazer_2007}. However, developing an \ac{OAC} scheme is not a trivial task due to the multipath channel, power misalignment, and time-synchronization errors in practice. Also, the \ac{CSI}  needs to be available at the \acp{ED} or the \ac{ES} with  state-of-the-art solutions. In this study, we propose an \ac{OAC} scheme to address these issues.

In the literature, various \ac{OAC} schemes are proposed for \ac{FEEL}. In \cite{Guangxu_2020}, analog modulation over \ac{OFDM} is investigated for \ac{BAA}. Particularly, it is proposed to modulate  the \ac{OFDM} subcarriers with the model parameters at the \acp{ED}. To overcome the impact of the multipath channel on the transmitted signals, the symbols on the \ac{OFDM} subcarriers are multiplied with the inverse of the channel coefficients and  the subcarriers that fade are excluded from the transmissions, which is known as {\em \ac{TCI}} in the literature. In \cite{sery2020overtheair}, an additional time-varying precoder is applied along with \ac{TCI} to facilitate the aggregation. 
In \cite{Amiri_2020}, it is proposed to sparsify the gradient estimates and project the resultant sparse vector into a low-dimensional vector to reduce the bandwidth. The compressed data is transmitted with \ac{BAA}.
In \cite{Guangxu_2021}, \ac{OBDA} is proposed  to facilitate the implementation of \ac{FEEL} for a practical wireless system. In this method,  considering distributed training by \ac{MV} with the  \ac{signSGD}~\cite{Bernstein_2018}, the \acp{ED} transmit \ac{QPSK} symbols  over \ac{OFDM} subcarriers along with \ac{TCI}, where the real and imaginary parts of the \ac{QPSK} symbols are formed by using the signs of the stochastic gradients, i.e., votes. At the \ac{ES}, 
the signs of the real and imaginary components of the superposed received symbols on each subcarrier
 are calculated to obtain the \ac{MV} for the sign of each gradient. 
However, the \acp{ED} still need the \ac{CSI} for \ac{TCI} as in \ac{BAA} for \ac{OAC}.  
In \cite{Yang_2020} and \cite{Amiria_2021}, blind \acp{ED} are considered. However, it is assumed that the \ac{CSI} for each \ac{ED} is available at the \ac{ES}. The impact of the channel on \ac{OAC} is mitigated through beamforming with a large number of antennas. 

In this study,  we investigate an \ac{OAC} method based on non-coherent detection to achieve \ac{FEEL} without using \ac{CSI} at the \acp{ED} and the \ac{ES}.  Inspired by the \ac{MV} with \ac{signSGD}~\cite{Bernstein_2018}, we use orthogonal resources, i.e., multiple subcarriers and/or \ac{OFDM} symbols, to transmit the signs of local stochastic gradients.
Hence, the votes from different \acp{ED} accumulate on the orthogonal resources non-coherently in fading channel with the proposed scheme. The \ac{ES} then obtains the \ac{MV} with an energy detector. 
Considering the randomness in the detected \acp{MV} due to the fading channel, path loss, and power control in the cell, we prove the convergence of learning in the presence of the proposed scheme for a non-convex loss function. We demonstrate that the proposed approach is robust against time-synchronization errors and power misalignment at the \ac{ES}. We also show that it can be used with well-known \ac{PMEPR} reduction techniques as it does not utilize the amplitude and the phase to encode the sign of local stochastic gradients. Finally, we evaluate the scheme by considering \ac{iid} data and non-\ac{iid} data where the data distribution is a function of the locations of \acp{ED}.


{\em Notation:} The complex  and real numbers are denoted by $\complexNumbers$ and  $\realNumbers$, respectively. 
$\expectationOperator[\cdot][]$ is the expectation of its  argument.
$\indicatorFunction[\cdot]$ is the indicator function and $\probability[\cdot]$ is the probability of its argument. The sign function is denoted by $\sign(\cdot)$ and results in $1$, $-1$, or $\pm1$ at random for a positive, a negative, or a zero-valued argument, respectively.

\section{System Model}
\label{sec:system}
\subsection{Scenario}
Consider a wireless network with $\numberOfEdgeDevices$ \acp{ED} that are connected to an \ac{ES}, where each \ac{ED} and the \ac{ES}  are equipped with single antennas.
We assume that the frequency synchronization in the network is done before the transmissions with a control mechanism as done in 3GPP \ac{4G} \ac{LTE} and/or \ac{5G} \ac{NR} with \ac{RACH} and/or \ac{PUCCH}  \cite{10.5555/3294673}.
In this study,  we consider the fact that the time synchronization  among the \acp{ED} is not ideal, and the maximum difference between the time of arrivals of the \acp{ED} signals at the \ac{ES} location is $\syncError$~seconds and it is equal to the reciprocal to the signal bandwidth.

In this study, the  power alignment at the \ac{ES} can be imperfect and the level of misalignment is controlled with a power control mechanism. We assume that the \ac{SNR} of an \ac{ED} at the \ac{ES} is $1/\noiseVariance$ at the reference distance $\referenceDistance$. We then set the received signal power of the $\indexED$th \ac{ED} at the \ac{ES} as
\begin{align}
\powerED[\indexED]=\left(\frac{{\distanceED[\indexED]}}{\referenceDistance}\right)^{-(\pathlossExponent-\powerControl)}~,
\end{align}
where $\distanceED[\indexED]$ is the link distance between the $\indexED$th \ac{ED} and the \ac{ES}, $\pathlossExponent$ is the path loss exponent, and $\powerControl\in [0,\pathlossExponent]$ is a coefficient that determines the amount of the path loss compensated. While $\powerControl=0$ means that there is no power control in the network, $\powerControl=\pathlossExponent$ leads to a system with perfect power alignment at the \ac{ES}. We define the effective path loss exponent $\effectivePathLossExponent$  as $\effectivePathLossExponent\triangleq\pathlossExponent-\powerControl$.

In this study, we assume that the \acp{ED} are deployed in a cell, where the cell radius is $\cellRadius$~meters and the minimum distance between the \ac{ES} and the \acp{ED} is $\minimumDistance$~meters for $\minimumDistance\ge\referenceDistance$. It is worth emphasizing that we do not consider the impact of multiple cells (e.g., inter-cell interference) or a more complicated large-scale channel model (e.g., shadowing) on learning in this work as our goal is to provide insights into the impact of power misalignment and the path loss on distributed learning with a tractable analysis.

\subsection{Signal Model}
In this study,   for \ac{OAC}, the \acp{ED} access the wireless channel  on the same time-frequency resources {\em simultaneously} with $\numberOfOFDMSymbols$ \ac{OFDM} symbols consisting of $\numberOfActiveSubcarriers$ active subcarriers.  
We assume that the \ac{CP} duration is larger than  $\syncError$ and the maximum-excess delays of the channel between the \ac{ES} and the \acp{ED}.
Considering independent frequency-selective channels between the \acp{ED} and the \ac{ES}, the superposed symbol on the $\indexSubcarrier$th subcarrier of the $\indexOFDMSymbol$th \ac{OFDM} symbol at the \ac{ES}  for the $\indexCommunicationRound$th communication  round of \ac{FEEL} can  be written as
\begin{align}
	\receivedSymbolAtSubcarrier[{\indexSubcarrier,\indexOFDMSymbol}] = \sum_{\indexED=1}^{\numberOfEdgeDevices}\sqrt{{\powerED[\indexED]}} \channelAtSubcarrier[\indexED,\indexSubcarrier,\indexOFDMSymbol]\transmittedSymbolAtSubcarrier[\indexED,\indexSubcarrier,\indexOFDMSymbol]+\noiseAtSubcarrier[\indexSubcarrier,\indexOFDMSymbol]~,
	\label{eq:symbolOnSubcarrier}
\end{align}
where  $\channelAtSubcarrier[\indexED,\indexSubcarrier,\indexOFDMSymbol]\in\complexNumbers$ is the channel coefficient between the \ac{ES} and the $\indexED$th \ac{ED}, $\transmittedSymbolAtSubcarrier[\indexED,\indexSubcarrier,\indexOFDMSymbol]\in\complexNumbers$ is the transmitted symbol from the $\indexED$th \ac{ED}, and $\noiseAtSubcarrier[\indexSubcarrier,\indexOFDMSymbol]$ is the symmetric \ac{AWGN} with zero mean and the variance $\noiseVariance$ on the $\indexSubcarrier$th subcarrier for $\indexSubcarrier\in\{0,1,\mydots,\numberOfActiveSubcarriers-1\}$ and $\indexOFDMSymbol\in\{0,1,\mydots,\numberOfOFDMSymbols-1\}$. 

We consider the fact that the time synchronization at the receiver may not be precise. To model this, we assume that the synchronization point where the \ac{DFT} starts can deviate by $\Nerror$ samples within the \ac{CP} window. Note that  the uncertainty of the synchronization point within the \ac{CP} window is often not an issue for traditional communications due to the channel estimation. However, it can cause a non-negligible impact on \ac{OAC}.

Let $\timeDomainOFDM[\timeVar]\in \complexNumbers$ be a baseband \ac{OFDM} symbol in continuous time for $\timeVar\in[0,\symbolDuration)$, where $\symbolDuration$ is the \ac{OFDM} symbol duration. 
We define the \ac{PMEPR} of an \ac{OFDM} symbol as  $\max_{\timeVar\in[0, \symbolDuration)}|\timeDomainOFDM[\timeVar]|^2/{\Ptransmit}
$, where $\Ptransmit=\expectationOperator[{|\timeDomainOFDM[\timeVar]|^2}][]$ is the mean-envelope power.

\subsection{Learning Model}
 Let $\dataset[\indexED]$ denote the local data containing labeled data samples at the $\indexED$th \ac{ED} as $\{(\sampleData[\indexSampleData], \sampleLabel[\indexSampleData] )\}\in\dataset[\indexED]$ for $\indexED=1,\mydots,\numberOfEdgeDevices$, where $\sampleData[\indexSampleData]$ and $\sampleLabel[\indexSampleData]$ are $\indexSampleData$th data sample and its associated label, respectively. The centralized learning problem can  be expressed as 
\begin{align}
	\modelParametersOptimal=\arg\min_{\modelParameters} \lossFunctionGlobal[\modelParameters]=\arg\min_{\modelParameters} \frac{1}{|\completeData|}\sum_{\forall(\sampleData[], \sampleLabel[] )\in\completeData} \lossFunctionSample[{\modelParameters,\sampleData[],\sampleLabel[]}]~,
	\label{eq:clp}
\end{align}
where $\completeData=\dataset[1]\cup\dataset[2]\cup\cdots\cup\dataset[K]$ and  $\lossFunctionSample[{\modelParameters,\sampleData[],\sampleLabel[]}]$ is the sample loss function that measures the labeling error for $(\sampleData[], \sampleLabel[])$ for the parameters $\modelParameters=[\modelParametersEle[1],\mydots,\modelParametersEle[\numberOfModelParameters]]^{\rm T}\in\realNumbers^{\numberOfModelParameters}$, and $\numberOfModelParameters$ is the number of parameters. With full-batch gradient descend, a local optimum point can be  obtained as
\begin{align}
\modelParametersAtIteration[\indexCommunicationRound+1] = \modelParametersAtIteration[\indexCommunicationRound] - \learningRate  \globalGradient[\indexCommunicationRound]~,
\end{align}
where $\learningRate$ is the learning rate and
\begin{align}
	\globalGradient[\indexCommunicationRound] =  \nabla \lossFunctionGlobal[{\modelParametersAtIteration[\indexCommunicationRound]}]
	= \frac{1}{|\completeData|}\sum_{\forall(\sampleData[], \sampleLabel[] )\in\completeData} \nabla 
	\lossFunctionSample[{\modelParametersAtIteration[\indexCommunicationRound],\sampleData[],\sampleLabel[]}]
	~,
	\label{eq:GlobalGradient}
\end{align}
where $\indexGradient$th element of the vector $\globalGradient[\indexCommunicationRound]$ is the gradient of $\lossFunctionGlobal[{\modelParametersAtIteration[\indexCommunicationRound]}]$ with respect to $\modelParametersAtIterationEle[\indexCommunicationRound][\indexGradient]$.

In  \cite{Bernstein_2018}, in the context of parallel processing, distributed training by \ac{MV} with \ac{signSGD} is investigated to solve \eqref{eq:clp}. In this method, for the $\indexCommunicationRound$th communication round, the $\indexED$th \ac{ED}\footnote{We refer  to the workers and parameter-server mentioned in \cite{Bernstein_2018}  as \acp{ED} and \ac{ES}, respectively, to describe distributed training by \ac{MV} with  \ac{signSGD}.} first calculates the local stochastic gradient as 
\begin{align}
	\localGradient[\indexED][\indexCommunicationRound] =  \nabla  \lossFunctionLocal[\indexED][{\modelParametersAtIteration[\indexCommunicationRound]}] 
	= \frac{1}{\batchSize} \sum_{\forall(\sampleData[\indexSampleData], \sampleLabel[\indexSampleData] )\in\datasetBatch[\indexED]} \nabla 
	\lossFunctionSample[{\modelParametersAtIteration[\indexCommunicationRound],\sampleData[\indexSampleData],\sampleLabel[\indexSampleData]}]
	~,
	\label{eq:LocalGradientEstimate}
\end{align}
where $\datasetBatch[\indexED]\subset\dataset[\indexED]$ is the selected data batch from the local data set and $\batchSize=|\datasetBatch[\indexED]|$ as the batch size. Instead of the actual values of local gradients, the \acp{ED} then send the signs of their local stochastic gradients, denoted as $\localGradientSign[\indexED][\indexCommunicationRound]$  for $\indexED=1,\mydots,\numberOfEdgeDevices$, to the \ac{ES}, where the $\indexGradient$th element of the vector $\localGradientSign[\indexED][\indexCommunicationRound]$ is
$\localGradientSignElement[\indexED,\indexGradient][\indexCommunicationRound]\triangleq\sign(\localGradientElement[\indexED,\indexGradient][\indexCommunicationRound])$. 
The \ac{ES}  obtains  the \ac{MV} for the $\indexGradient$th gradient as
 \begin{align}
 	\majorityVoteEle[\indexCommunicationRound][\indexGradient]\triangleq\sign\left(\sum_{\indexED=1}^{\numberOfEdgeDevices} \localGradientSignElement[\indexED,\indexGradient][\indexCommunicationRound]\right)~.
 	\label{eq:majorityVote}
 \end{align} 
Subsequently, the \ac{ES} pushes $\majorityVote[\indexCommunicationRound]=[\majorityVoteEle[\indexCommunicationRound][1],\mydots,\majorityVoteEle[\indexCommunicationRound][\numberOfModelParameters]]^{\rm T}$ to the \acp{ED}  and the models at the \acp{ED} are updated as
\begin{align}
\modelParametersAtIteration[\indexCommunicationRound+1] = \modelParametersAtIteration[\indexCommunicationRound] - \learningRate  \majorityVote[\indexCommunicationRound]~.
\label{eq:MVsignSGD}
\end{align}
This procedure is repeated consecutively until a predetermined convergence criterion is achieved. 

 For \ac{FEEL}, the optimization problem can also be expressed as \eqref{eq:clp} in a scenario where the local data samples and their labels are not available at the \ac{ES} and the link between an \ac{ED} and the \ac{ES} experiences independent frequency-selective fading channel. To solve \eqref{eq:clp} under these constraints, in this study, we adopt the same procedure summarized for the distributed training by the \ac{MV}.  With the motivations of eliminating the latency caused by orthogonal multiple access and enabling distributed training in {\em mobile} wireless networks, we propose a simple-but-effective \ac{OAC} scheme to detect the \ac{MV} in fading channel without using \ac{CSI} at the \acp{ED} and the \ac{ES}.

\section{FSK-Based Majority Vote}
\label{sec:fskMV}

\subsection{Edge Device - Transmitter}
With the proposed \ac{OAC} scheme, the \acp{ED} perform a low-complexity operation to transmit the signs of the  gradients given in \eqref{eq:LocalGradientEstimate}:
Let $\mappingFunction$ be a bijective function that maps $\indexGradient\in\{1,2,\mydots,\numberOfModelParameters\}$ to the distinct pairs $(\voteInTime[+],\voteInFrequency[+])$ and $(\voteInTime[-],\voteInFrequency[-])$ for $\voteInTime[+],\voteInTime[-]\in\{0,1,\mydots,\numberOfOFDMSymbols-1\})$ and $\voteInFrequency[+],\voteInFrequency[-]\in\{0,1,\mydots,\numberOfActiveSubcarriers-1\}$. 
 Based on the value of $\localGradientSignElement[\indexED,\indexGradient][\indexCommunicationRound]$, at the $\indexCommunicationRound$th communication round,  the $\indexED$th \ac{ED} calculates the symbol $\transmittedSymbolAtSubcarrier[\indexED,{\voteInFrequency[+]},{\voteInTime[+]}]$ and $\transmittedSymbolAtSubcarrier[\indexED,{\voteInFrequency[-]},{\voteInTime[-]}]$, $\forall\indexGradient$, as\begin{align}
	\transmittedSymbolAtSubcarrier[\indexED,{\voteInFrequency[+]},{\voteInTime[+]}]=\begin{cases}
			\sqrt{\symbolEnergy}\times \randomSymbolAtSubcarrier[\indexED,\indexGradient]& \localGradientSignElement[\indexED,\indexGradient][\indexCommunicationRound] =1,\\
			0,& \localGradientSignElement[\indexED,\indexGradient][\indexCommunicationRound] =-1\\
	\end{cases}~,
\label{eq:symbolOne}
\end{align}
and
\begin{align}
	\transmittedSymbolAtSubcarrier[\indexED,{\voteInFrequency[-]},{\voteInTime[-]}]=\begin{cases}
		0, & \localGradientSignElement[\indexED,\indexGradient][\indexCommunicationRound] =1\\
		\sqrt{\symbolEnergy}\times \randomSymbolAtSubcarrier[\indexED,\indexGradient],& \localGradientSignElement[\indexED,\indexGradient][\indexCommunicationRound] =-1\\
	\end{cases}~,
\label{eq:symbolTwo}
\end{align}
respectively, where $\symbolEnergy=2$ is a factor to normalize the symbol energy and $\randomSymbolAtSubcarrier[\indexED,\indexGradient]$ is a randomization symbol on the unit circle. Therefore, to indicate the sign of a local stochastic gradient, our scheme dedicates two subcarriers with \eqref{eq:symbolOne} and \eqref{eq:symbolTwo},  as opposed to modulating the phase of a subcarrier as done in \ac{OBDA}. Also, we do not use \ac{TCI} to compensate the impact of multipath channel on transmitted symbols as our goal is to exploit the energy accumulation on two different subcarriers to detect the \ac{MV} with a non-coherent detector. 

As a special case of $\mappingFunction$, if $\voteInTime[-]=\voteInTime[+]$ and $\voteInFrequency[-]=\voteInFrequency[+]+1$ hold for all $\indexGradient$, the adjacent subcarriers of $\voteInTime[+]$th \ac{OFDM} symbol forms the options for a vote, which corresponds to \ac{FSK} over \ac{OFDM} subcarriers.  In this case,
the $\indexED$th \ac{ED}'s vote for the $\indexGradient$th gradient becomes independent from its choice
since the adjacent subcarriers are likely to experience similar channel conditions, i.e., $\channelAtSubcarrier[{\indexED,\voteInFrequency[+]}]\approxeq\channelAtSubcarrier[{\indexED,\voteInFrequency[+]+1}]$. %
 We refer to the \ac{MV} calculation with the proposed scheme under this specific mapping as {\ac{FSK-MV}} in this study.  
 
 After the calculations of $\transmittedSymbolAtSubcarrier[\indexED,{\voteInFrequency[+]},{\voteInTime[+]}]$ and $\transmittedSymbolAtSubcarrier[\indexED,{\voteInFrequency[-]},{\voteInTime[-]}]$ for all $\indexGradient$ and $\indexED$, the \acp{ED} calculate the \ac{OFDM} symbols and transmit them based on the discussions in Section~\ref{sec:system}.

\subsection{Edge Server - Receiver}
\begin{figure*}[t]
	\centering
	{\includegraphics[width =6in]{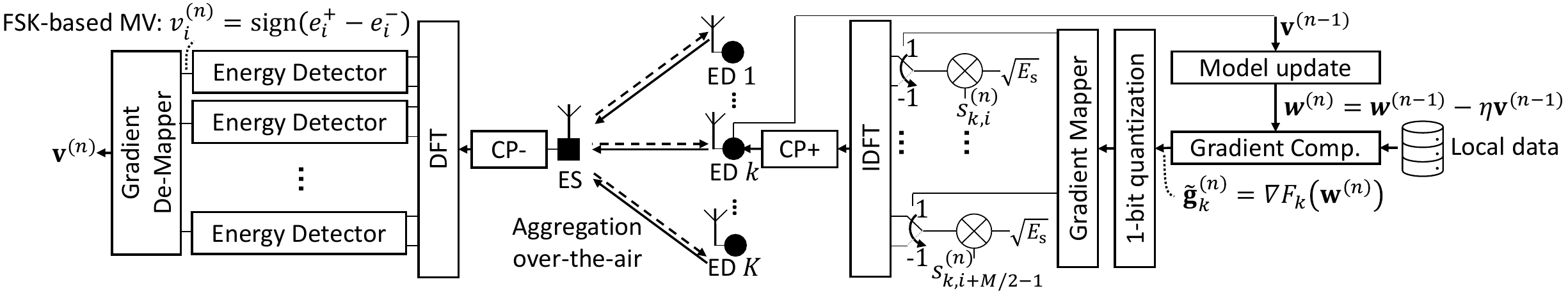}
	} 
	\caption{Transmitter and receiver diagrams  for implementing distributed learning  over a wireless network with \ac{FSK-MV}.}
	\label{fig:feelBlockDiagram}
\end{figure*}
The receiver at the \ac{ES} observes the superposed symbols at all subcarriers as expressed in \eqref{eq:symbolOnSubcarrier}. By using the mapping function  $\mappingFunction$, the superposed symbols for a given $\indexGradient$ can be shown  as
\begin{align}
\receivedSymbolAtSubcarrier[{\voteInFrequency[+]},{\voteInTime[+]}]
	=\sqrt{\symbolEnergy}\sum_{\substack{\forall\indexED, \localGradientSignElement[\indexED,\indexGradient][\indexCommunicationRound]=1}} \sqrt{{\powerED[\indexED]}} \channelAtSubcarrier[\indexED,{\voteInFrequency[+]},{\voteInTime[+]}]\randomSymbolAtSubcarrier[\indexED,\indexGradient] +\noiseAtSubcarrier[{\voteInFrequency[+]},{\voteInTime[+]}]~,
	\label{eq:superposedOne}
\end{align}
and 
\begin{align}
\hspace{-2mm}\receivedSymbolAtSubcarrier[{\voteInFrequency[-]},{\voteInTime[-]}]=\sqrt{\symbolEnergy}\sum_{\substack{\forall\indexED, \localGradientSignElement[\indexED,\indexGradient][\indexCommunicationRound]=-1}} \sqrt{{\powerED[\indexED]}} \channelAtSubcarrier[\indexED,{\voteInFrequency[-]},{\voteInTime[-]}]\randomSymbolAtSubcarrier[\indexED,\indexGradient] +\noiseAtSubcarrier[{\voteInFrequency[-]},{\voteInTime[-]}].
	\label{eq:superposedTwo}
\end{align}
respectively. The receiver at the \ac{ES} detects the \ac{MV}  for the $\indexGradient$th gradient with an energy detector as
\begin{align}
	\majorityVoteEle[\indexCommunicationRound][\indexGradient] = \signNormal[{\deltaVectorAtIteration[\indexCommunicationRound][\indexGradient]}]~,
	\label{eq:detector}
\end{align}
where $\deltaVectorAtIteration[\indexCommunicationRound][\indexGradient]\triangleq{\metricForFirst[\indexGradient]-\metricForSecond[\indexGradient]}$  for
$
\metricForFirst[\indexGradient]\triangleq  |\receivedSymbolAtSubcarrier[{\voteInFrequency[+]},{\voteInTime[+]}]|_2^2
$
and
$
\metricForSecond[\indexGradient]\triangleq  |\receivedSymbolAtSubcarrier[{\voteInFrequency[+]},{\voteInTime[+]}]|_2^2$, $\forall\indexGradient$. It is worth mentioning that we do not use any method to resolve the interference  in \eqref{eq:superposedOne} and \eqref{eq:superposedTwo}  among the \acp{ED} as we are not interested in the sign of a local gradients. On the contrary, we exploit the interference for aggregation and compare the amount of energy on two different subcarriers to detect the \ac{MV} in \eqref{eq:detector}. The transmitter and receiver block diagrams are provided in \figurename~\ref{fig:feelBlockDiagram}, based on the aforementioned discussions.

The proposed scheme leads to a fundamentally different training strategy  since 
it determines the correct \ac{MV} in \eqref{eq:majorityVote} {\em probabilistically} by comparing ${\metricForFirst[\indexGradient]}$ and ${\metricForSecond[\indexGradient]}$. 
To elaborate this, assume that  the multipath channels between the \ac{ES} and the \acp{ED} are independent. Let $\numberOFEDsForOptionOne$ and $\numberOFEDsForOptionSecond=\numberOfEdgeDevices-\numberOFEDsForOptionOne$ be the number of \acp{ED} that vote for $1$ and $-1$ for the $\indexGradient$th gradient, respectively. 
\begin{lemma} \rm 	$\expectationOperator[{\metricForFirst[\indexGradient]}][]$ and $\expectationOperator[{\metricForSecond[\indexGradient]}][]$ can be calculated as
	\begin{align}
		\meanOptionOne\triangleq\expectationOperator[{\metricForFirst[\indexGradient]}][]
		=\symbolEnergy\numberOFEDsForOptionOne\largeScaleImpactOnLearning+\noiseVariance~,
		\label{eq:energyFirst}
	\end{align}
	and 
	\begin{align}
		\meanOptionTwo\triangleq\expectationOperator[{\metricForSecond[\indexGradient]}][]
		=&\symbolEnergy\numberOFEDsForOptionSecond\largeScaleImpactOnLearning+\noiseVariance~,
		\label{eq:energySecond}
	\end{align}
	respectively, where
	\begin{align}
		\largeScaleImpactOnLearning \triangleq \begin{cases}
			\frac{2\referenceDistance^{\effectivePathLossExponent}}{\cellRadius^2-\minimumDistance^2} \frac{\minimumDistance^{2-\effectivePathLossExponent}-\cellRadius^{2-\effectivePathLossExponent}}{\effectivePathLossExponent-2}~, & \effectivePathLossExponent\neq2\\
			\frac{2\referenceDistance^{\effectivePathLossExponent}}{\cellRadius^2-\minimumDistance^2}\ln{\frac{\cellRadius}{\minimumDistance}}~, & \effectivePathLossExponent=2\\
		\end{cases}~.
	\label{eq:pathlossimpact}
	\end{align}
	\label{lemma:exp}
\end{lemma}
\begin{IEEEproof}
Since \eqref{eq:superposedOne} is a weighted summation of independent complex Gaussian random variables with zero mean and unit variance (i.e., channel coefficients), $\receivedSymbolAtSubcarrier[{\voteInFrequency[+]},{\voteInTime[+]}]$ is a zero mean random variable, where its variance is
\begin{align}
 \meanOptionOne= \expectationOperator[{\metricForFirst[\indexGradient]}][]=&\expectationOperator[{|\receivedSymbolAtSubcarrier[{\voteInFrequency[+]},{\voteInTime[+]}]|_2^2}][]= \expectationOperator[{\symbolEnergy\sum_{\substack{\localGradientSignElement[\forall\indexED,\indexGradient][\indexCommunicationRound]=1}}\left(\frac{{\distanceED[\indexED]}}{\referenceDistance}\right)^{-\effectivePathLossExponent}+\noiseVariance}][]
 \nonumber\\
 =&\symbolEnergy\numberOFEDsForOptionOne\expectationOperator[{\left(\frac{{\distanceED[\indexED]}}{\referenceDistance}\right)^{-\effectivePathLossExponent}}][]+\noiseVariance~.
 \label{eq:mean}
\end{align}
To calculate \eqref{eq:mean}, we need to calculate the expected value of $\arandomvar=\distanceED[]^{-\effectivePathLossExponent}$. Assuming that the \acp{ED} are localized uniformly within the cell, the link distance distribution can be expressed as
\begin{align}
	f(\distanceED[])=\frac{2\distanceED[]}{\cellRadius^2-\minimumDistance^2}~.
	\label{eq:linkDistance}
\end{align}
Hence, the distribution of $\arandomvar$ can obtained as
\begin{align}
f(\arandomvar)=\frac{	f(\distanceED[])}{|\frac{d\arandomvar}{d\distanceED[]}|}\bigg|_{\distanceED[]=\arandomvar^{-\frac{1}{\effectivePathLossExponent}}}=\frac{2\arandomvar^{-\frac{\effectivePathLossExponent+2}{\effectivePathLossExponent}}}{(\cellRadius^2-\minimumDistance^2)\effectivePathLossExponent}.
\label{eq:fdist}
\end{align}
By using \eqref{eq:fdist}, the expected value of $\arandomvar$ can be calculated as \eqref{eq:pathlossimpact}. The same analysis can be done for $\meanOptionTwo$.
\end{IEEEproof}

Based on Lemma~\ref{lemma:exp}, \eqref{eq:detector} is likely to obtain the correct \ac{MV} because $\meanOptionOne$ and $\meanOptionTwo$ are linear functions of $\numberOFEDsForOptionOne$ and $\numberOFEDsForOptionSecond$, respectively. However, the detection performance depends on the parameter $\largeScaleImpactOnLearning\in[0,1]$ that captures the impacts of power control, path loss, and cell size on ${\metricForFirst[\indexGradient]}$ and ${\metricForSecond[\indexGradient]}$.
In \figurename~\ref{fig:lambda}, we plot $\largeScaleImpactOnLearning$ for different cell sizes for a given $\effectivePathLossExponent$. For a better power control or a smaller cell size, the parameter  $\largeScaleImpactOnLearning$ increases to $1$, which implies a better detection performance under noise. On the other hand, the \ac{MV} is not deterministic for $\noiseVariance=0$. Hence, the convergence for a non-convex loss function $\lossFunctionGlobal[\modelParameters]$  needs to be shown to justify if the proposed scheme is suitable for \ac{FEEL}. 

\subsection{Convergence in Fading Channel}
\label{ssec:why}
We consider several standard assumptions made in the literature for the convergence analysis \cite{Bernstein_2018, Guangxu_2021}:
\begin{assumption}[Bounded loss function]
	\rm 
	$\lossFunctionGlobal[\modelParameters]\ge \lossFunctionGlobalMinimum$, $\forall\modelParameters$. 
\end{assumption}
\begin{assumption}[Smoothness] 
	\rm 
	Let $\globalGradientNoIndex$ be the gradient of $\lossFunctionGlobal[\modelParameters]$ evaluated at $\modelParameters$. For all $\modelParameters$ and $\modelParameters'$, the expression given by
	\begin{align}
		\left| \lossFunctionGlobal[\modelParameters'] - (\lossFunctionGlobal[\modelParameters]-\globalGradientNoIndex^{\rm T}(\modelParameters'-\modelParameters)) \right| \le \frac{1}{2}\sum_{\indexGradient=1}^{\numberOfModelParameters} \nonnegativeConstantsEle[\indexGradient](\modelParametersEle[\indexGradient]'-\modelParametersEle[\indexGradient])^2~,
		\nonumber
	\end{align}	
	holds for a non-negative constant vector
	$\nonnegativeConstants=[\nonnegativeConstantsEle[1],\mydots,\nonnegativeConstantsEle[\numberOfModelParameters]]^{\rm T}$.
\end{assumption}
\begin{assumption}[Variance bound]
	\rm The stochastic  gradient estimates $\{\localGradientNoIndex[\indexED]=[\localGradientNoIndexElement[\indexED,1],\mydots,\localGradientNoIndexElement[\indexED,\numberOfModelParameters]]^{\rm T}=\nabla \lossFunctionLocal[\indexED][{\modelParametersAtIteration[\indexCommunicationRound]}]\} $, $\forall\indexED$, are independent and  unbiased estimates of $\globalGradientNoIndex=[\globalGradientElementNoIndex[1],\mydots,\globalGradientElementNoIndex[\numberOfModelParameters]]^{\rm T}=\nabla\lossFunctionGlobal[{\modelParameters}]$ with a coordinate bounded variance, i.e.,
	\begin{align}
		\expectationOperator[{\localGradientNoIndex[\indexED]}][]&=\globalGradientNoIndex,~\forall\indexED,\\	\expectationOperator[{(\localGradientNoIndexElement[\indexED,\indexGradient]-\globalGradientElementNoIndex[\indexGradient])^2}][]&\le\varianceBoundEle[\indexGradient]^2/\batchSize,~\forall\indexED,\indexGradient,
	\end{align}
	where  $\varianceBound = [\varianceBoundEle[1],\mydots,\varianceBoundEle[\numberOfModelParameters]]^{\rm T}$ is a non-negative constant  vector.
\end{assumption}
\begin{assumption}[Unimodal, symmetric gradient noise]
	\rm
	For any given $\modelParameters$, the elements of the vector $\localGradientNoIndex[\indexED]$, $\forall\indexED$, has a unimodal distribution that is also symmetric around its mean.
\end{assumption}

\begin{figure}[t]
	\centering
	{\includegraphics[width =3.5in]{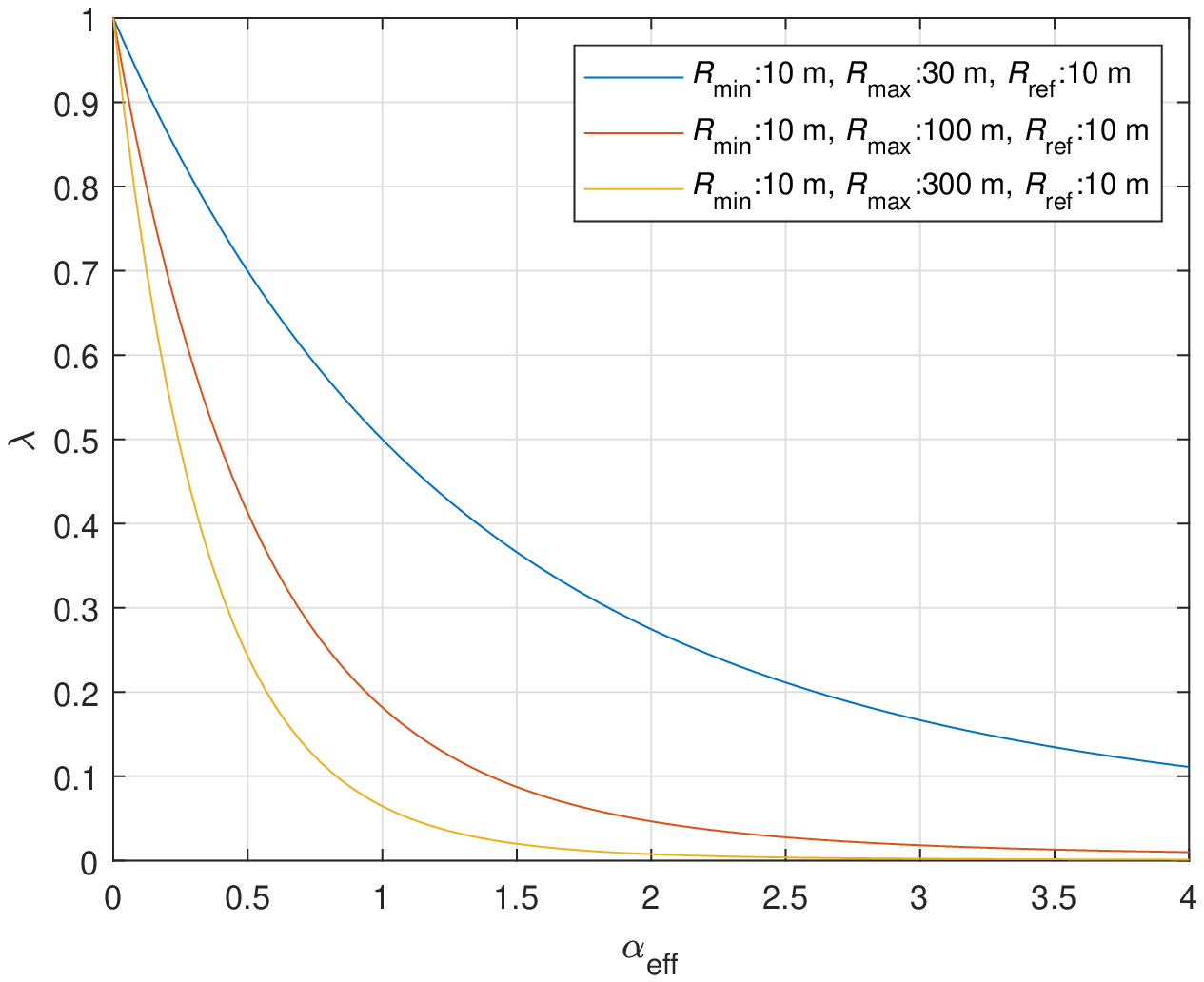}
	} 
	\caption{Impact of cell size and the effective path loss exponent on  $\largeScaleImpactOnLearning$.}
	\label{fig:lambda}
\end{figure}
We also assume that the parameters	${\metricForFirst[\indexGradient]}$ and ${\metricForSecond[\indexGradient]}$ are exponential random variables, where their means are ${\meanOptionOne}$ and ${\meanOptionTwo}$, respectively.
This assumption holds true when the power control is ideal under \ac{iid} Rayleigh fading. It is a weak assumption under imperfect power control due to the central limit theorem.

By extending our theorem in \cite{sahinCCNC_2022_submit} with the considerations of path loss, power control, and cell size, the convergence rate in the presence of \ac{FSK-MV} can obtained as follows:
\begin{theorem}
	\rm For $\batchSize=\communicationRounds/\brachSizeRelativeToRounds$ and $\learningRate=1/\sqrt{\norm{\nonnegativeConstants}_1\batchSize}$, the convergence rate of the distributed training by the \ac{MV} based on \ac{FSK} in fading channel is
	\begin{align}
		\expectationOperator[\frac{1}{\communicationRounds}\sum_{\indexCommunicationRound=0}^{\communicationRounds-1} \norm{\globalGradient[\indexCommunicationRound]}_1][]\nonumber\le\frac{1}{\sqrt{\communicationRounds}}&\bigg( \coefficientOne\sqrt{\norm{\nonnegativeConstants}_1}\left(	\lossFunctionGlobal[{\modelParametersAtIteration[0]}]- \lossFunctionGlobalMinimum+\frac{\brachSizeRelativeToRounds}{2}\right)\nonumber\\&~~+\frac{2\sqrt{2}}{3}\sqrt{\brachSizeRelativeToRounds}\norm{\varianceBound}_1\bigg)~,
		\label{eq:convergence}
	\end{align}
	where $\brachSizeRelativeToRounds$ is a positive integer,  $\coefficientOne=(1+\frac{2}{\effectiveSNR\numberOfEdgeDevices})\frac{1}{\sqrt{\brachSizeRelativeToRounds}}$  for $\effectiveSNR\triangleq\frac{\symbolEnergy\largeScaleImpactOnLearning}{\noiseVariance} $, and $\largeScaleImpactOnLearning\in[0,1]$ given in \eqref{eq:pathlossimpact} is a parameter that  captures the parameters related to the path loss, power control, and cell size.
	\label{th:convergence}
\end{theorem}
The proof of Theorem~\ref{th:convergence} is given in the appendix.

Based on Theorem~\ref{th:convergence}, we can infer the followings: 1) For a larger \ac{SNR} (i.e., a larger $1/\noiseVariance$)  and a large number of \acp{ED} (i.e., a larger $\numberOfEdgeDevices$), the convergence rate with \ac{FSK-MV} in fading channel improves since $\coefficientOne$ decreases. 2) The power control results in a better convergence rate since $\lambda$ increases with a lower $\effectivePathLossExponent$. 3) Another way of improving the convergence rate is to reduce to cell size, yielding a large $\lambda$ as illustrated in \figurename~\ref{fig:lambda}. However, this indicates a practical limitation of a single-cell \ac{FEEL}: The number of \acp{ED} may be smaller for a smaller cell. However, the power control becomes a harder task for a larger cell. 4) Finally, under ideal power control, the convergence rate becomes similar to the one with \ac{signSGD} in an ideal channel \cite[Theorem~1]{Bernstein_2018} asymptotically.

\subsection{Comparisons}
\label{subsec:comp}
\subsubsection{Robustness against Time-Varying Fading Channel} As opposed to the approaches in \cite{Guangxu_2020} and \cite{Guangxu_2021}, the proposed scheme does not utilize the \ac{CSI} for \ac{TCI} at the \acp{ED}. Hence, it is compatible with time-varying channels (e.g., mobile networks \cite{Zeng_2020}) and does not lose gradient information due to \ac{TCI}. 
As a trade-off, it quadruples the number of time-frequency resources for \ac{OAC} as compared to  \ac{OBDA} in \cite{Guangxu_2021}. As compared to the approaches in \cite{Yang_2020} and \cite{Amiria_2021}, the proposed scheme also does not require \ac{CSI} at the \ac{ES} or multiple antennas. 

\subsubsection{Robustness against Time-Synchronization Errors} 
As demonstrated in Section~\ref{sec:numerical}, the proposed scheme provides immunity against the time-synchronization errors. This is because the timing misalignment among the \acp{ED} or the uncertainty on the receiver synchronization within the \ac{CP} window cause phase rotations in the frequency domain and \ac{FSK-MV} does not encode information on the amplitude or phase. Also, the proposed scheme does not use any channel-related information at the \acp{ED} and the \ac{ES}. Hence, \ac{FSK-MV} is more robust against time-synchronization errors as compared to \ac{OBDA}.

\subsubsection{Robustness against Power-Amplifier Non-linearity} 
The proposed scheme separates the options for voting over two different resources identified in time and frequency. Hence, it allows one to choose $\randomSymbolAtSubcarrier[\indexED,\indexGradient]$ based on specific purposes. In this study, we use random \ac{QPSK} symbol to reduce \ac{PMEPR} by decreasing the correlation in the frequency domain \cite{Jawhar_2019}. 
\ac{OBDA} is not investigated in terms of \ac{PMEPR} in the literature. As shown in Section~\ref{sec:numerical}, \ac{OBDA} can suffer from high \ac{PMEPR}, while the proposed scheme reduces \ac{PMEPR} with a simple randomization technique. Also, \ac{FSK-MV} does not require a long transmission power constraint as in introduced for \ac{OBDA} \cite[Eq. 9 and Eq. 10]{Guangxu_2021} since the $\ell_2$-norm of the \ac{OFDM} symbols do not change  as a function of \ac{CSI}  with \ac{FSK-MV}.

\section{Numerical Results}
\label{sec:numerical}

\begin{table}[t]
	\centering
	\caption{Neural network at the EDs.}
	\begin{tabular}{l|l}
		Layer               				& Learnables \\ \hline\hline
		Input ($28\times28\times1$ images) & N/A\\\hline
		Convolution 2D ($5\times5$, $20$ filters)   	& \begin{tabular}[c]{@{}l@{}}Weights: $5\times5\times1\times20$\\ Bias: $1\times1\times20$ \end{tabular} \\\hline
		Batchnorm & \begin{tabular}[c]{@{}l@{}}Offset: $1\times1\times20$\\ Scale: $1\times1\times20$ \end{tabular} \\ \hline
		ReLU &  N/A\\\hline
		Convolution 2D ($3\times3$, $20$ filters)   	& \begin{tabular}[c]{@{}l@{}}Weights: $3\times3\times20\times20$\\ Bias: $1\times1\times20$ \end{tabular} \\\hline
		Batchnorm & \begin{tabular}[c]{@{}l@{}}Offset: $1\times1\times20$\\ Scale: $1\times1\times20$ \end{tabular} \\ \hline
		ReLU &  N/A\\		\hline
		Convolution 2D ($3\times3$, $20$ filters)    	& \begin{tabular}[c]{@{}l@{}}Weights: $3\times3\times20\times20$\\ Bias: $1\times1\times20$ \end{tabular} \\\hline
		Batchnorm & \begin{tabular}[c]{@{}l@{}}Offset: $1\times1\times20$\\ Scale: $1\times1\times20$ \end{tabular} \\ \hline
		ReLU &  N/A\\		\hline
		Fully-connected layer ($10$  outputs)     					&    \begin{tabular}[c]{@{}l@{}}Weights: $10\times11520$\\ Bias: $10\times1$ \end{tabular}        \\\hline
		Softmax &  N/A\\\hline 
	\end{tabular}
	\label{table:layout}
\end{table}
\begin{figure}[t]
	\centering
	\subfloat[IID data in the cell. All \acp{ED} have data samples for 10 different digits.]{\includegraphics[width =3.5in]{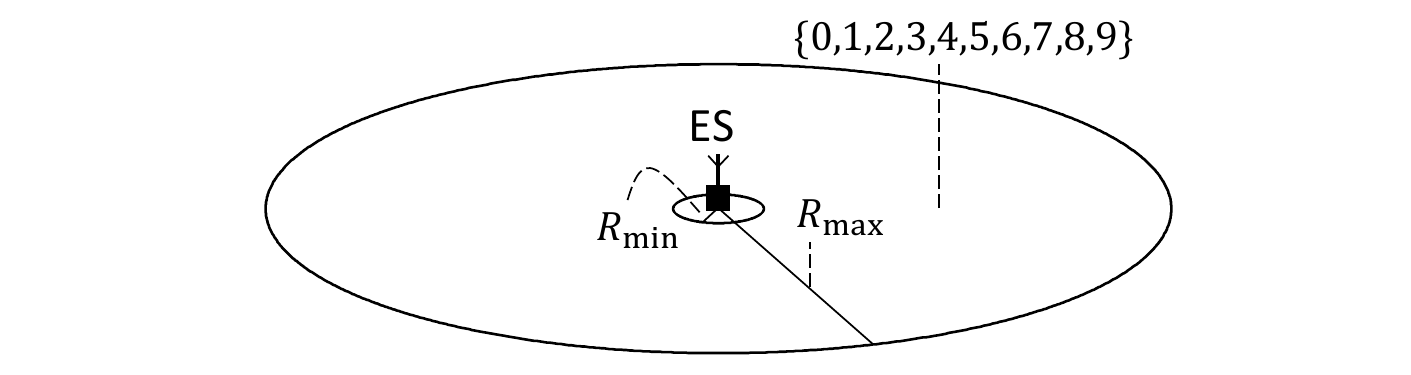}}\\
	\subfloat[Non-IID data in the cell. The available digits at the \acp{ED} change based on their locations in the cell. The  digits in an area are shown in the figure.]{\includegraphics[width =3.5in]{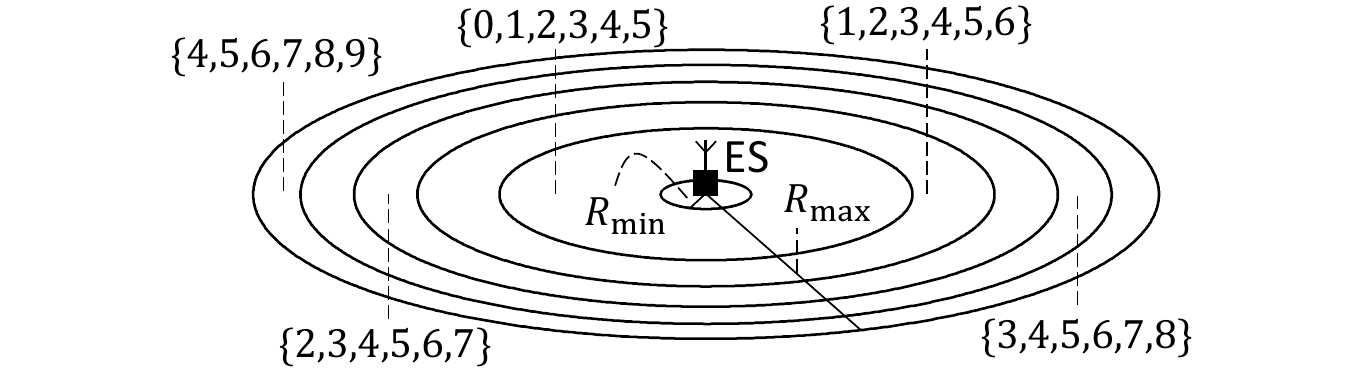}}
	\caption{IID versus non-IID data considered for the numerical analyses. The radius of the concentric circles are $\{10,45.6,63.7,77.7,89.6,100\}$ meters.}
	\label{fig:niid}
\end{figure}
\def\figuresize{3.0in}
\begin{figure*}[t]
	\centering
	\subfloat[IID data, ideal power control ($\effectivePathLossExponent=0$).]{\includegraphics[width =\figuresize]{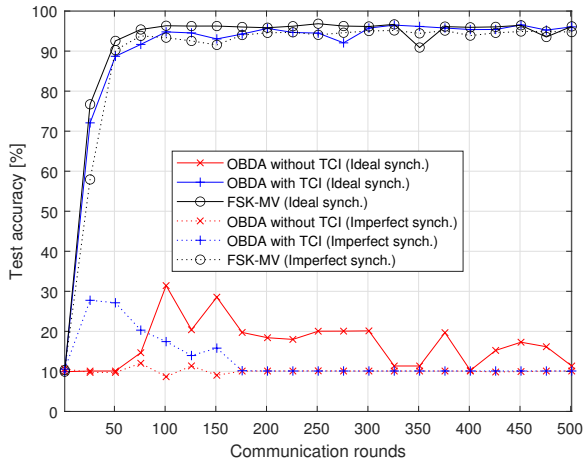}
		\label{subfig:acc_iid_aeff_zero}}~~~~~~~~~~~
	\subfloat[IID data, imperfect power control  ($\effectivePathLossExponent=2$).]{\includegraphics[width =\figuresize]{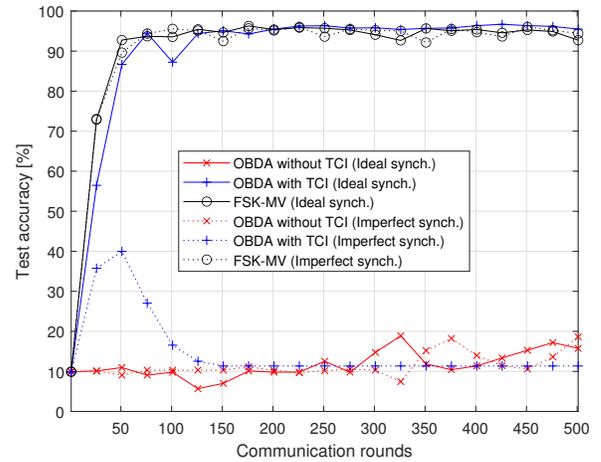}
		\label{subfig:acc_iid_aeff_two}}\\
	\subfloat[Non-IID data, ideal power control  ($\effectivePathLossExponent=0$).]{\includegraphics[width =\figuresize]{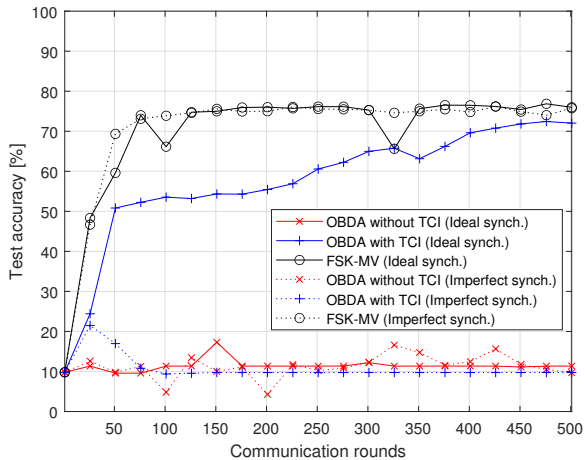}
		\label{subfig:acc_niid_aeff_zero}}~~~~~~~~~~~	
	\subfloat[Non-IID data, imperfect power control  ($\effectivePathLossExponent=2$).]{\includegraphics[width =\figuresize]{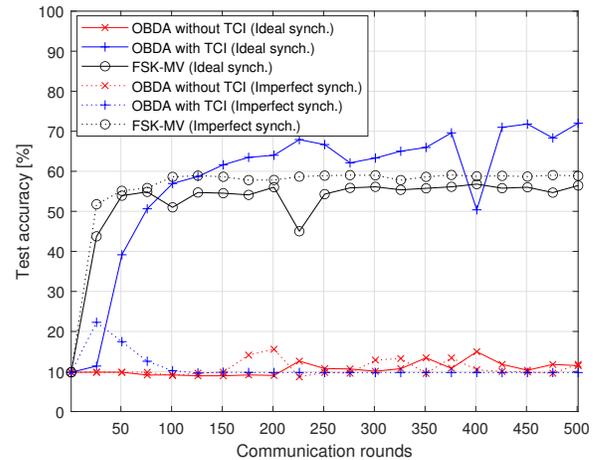}
		\label{subfig:acc_niid_aeff_two}}		
	\caption{Test accuracy versus communication rounds. \ac{FSK-MV} works without the  \ac{CSI} at the \acp{ED} and \ac{ES} and provide robustness against time-synchronization errors. The test accuracy reduces more for non-\ac{iid} when the power control is imperfect.}
	\label{fig:testAcc}
\end{figure*}
\begin{figure*}[t]
	\centering
	\subfloat[IID data, ideal power control ($\effectivePathLossExponent=0$).]{\includegraphics[width =\figuresize]{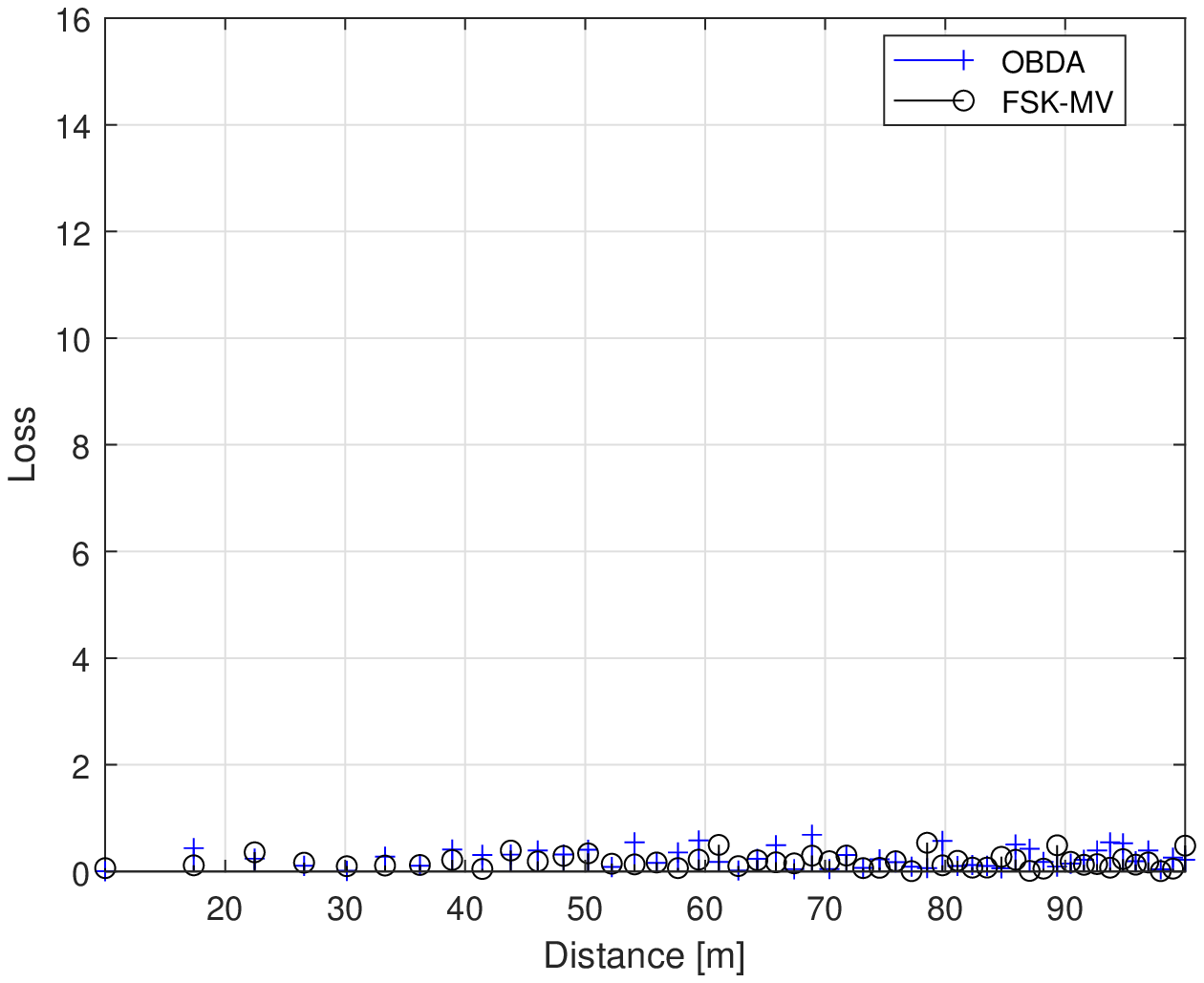}
		\label{subfig:ld_iid_aeff_zero}}~~~~~~~~~~~
	\subfloat[IID data, imperfect power control ($\effectivePathLossExponent=2$).]{\includegraphics[width =\figuresize]{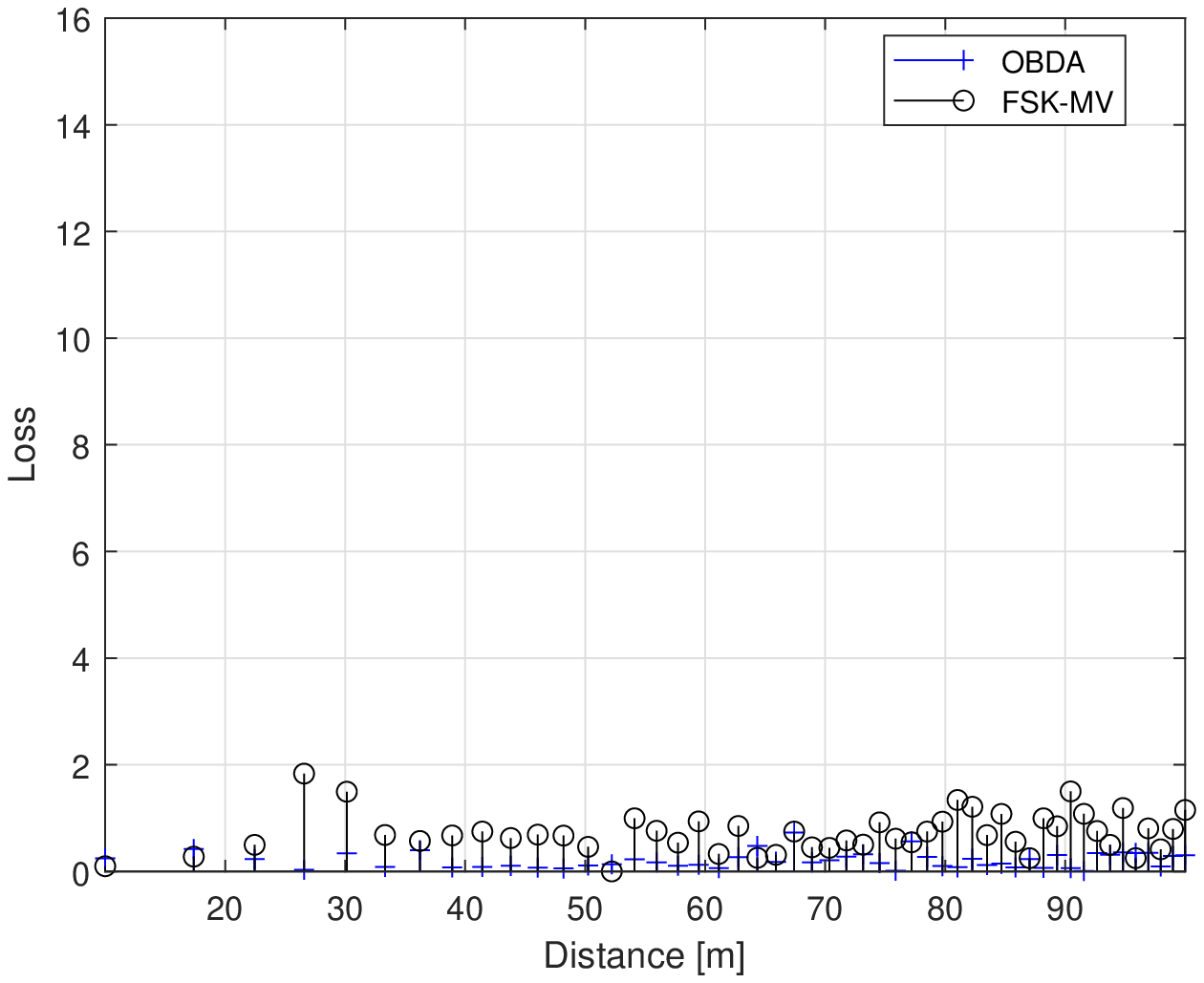}
		\label{subfig:ld_iid_aeff_two}}\\
	\subfloat[Non-IID data, ideal power control ($\effectivePathLossExponent=0$).]{\includegraphics[width =\figuresize]{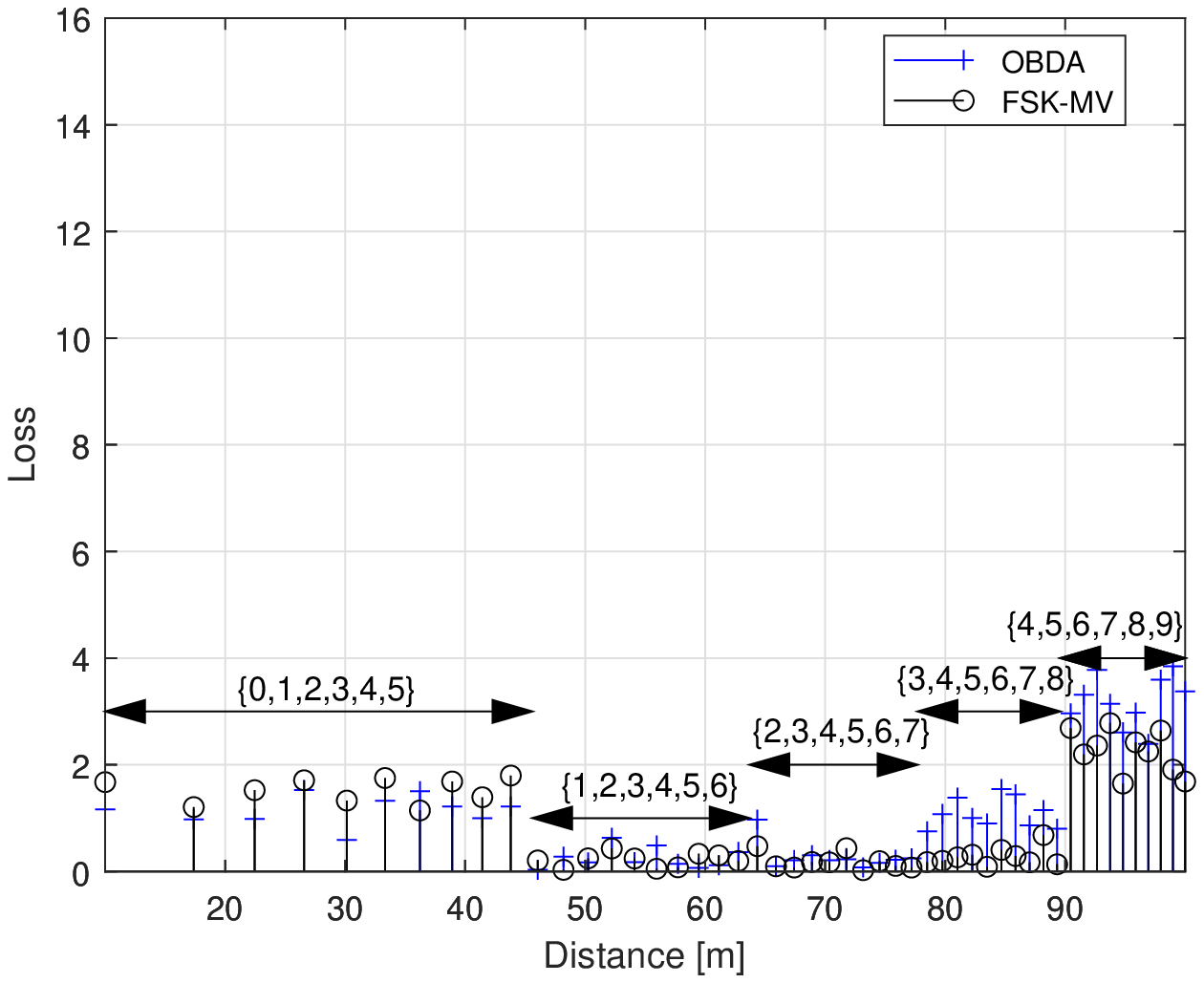}
		\label{subfig:ld_niid_aeff_zero}}~~~~~~~~~~~	
	\subfloat[Non-IID data, imperfect power control ($\effectivePathLossExponent=2$).]{\includegraphics[width =\figuresize]{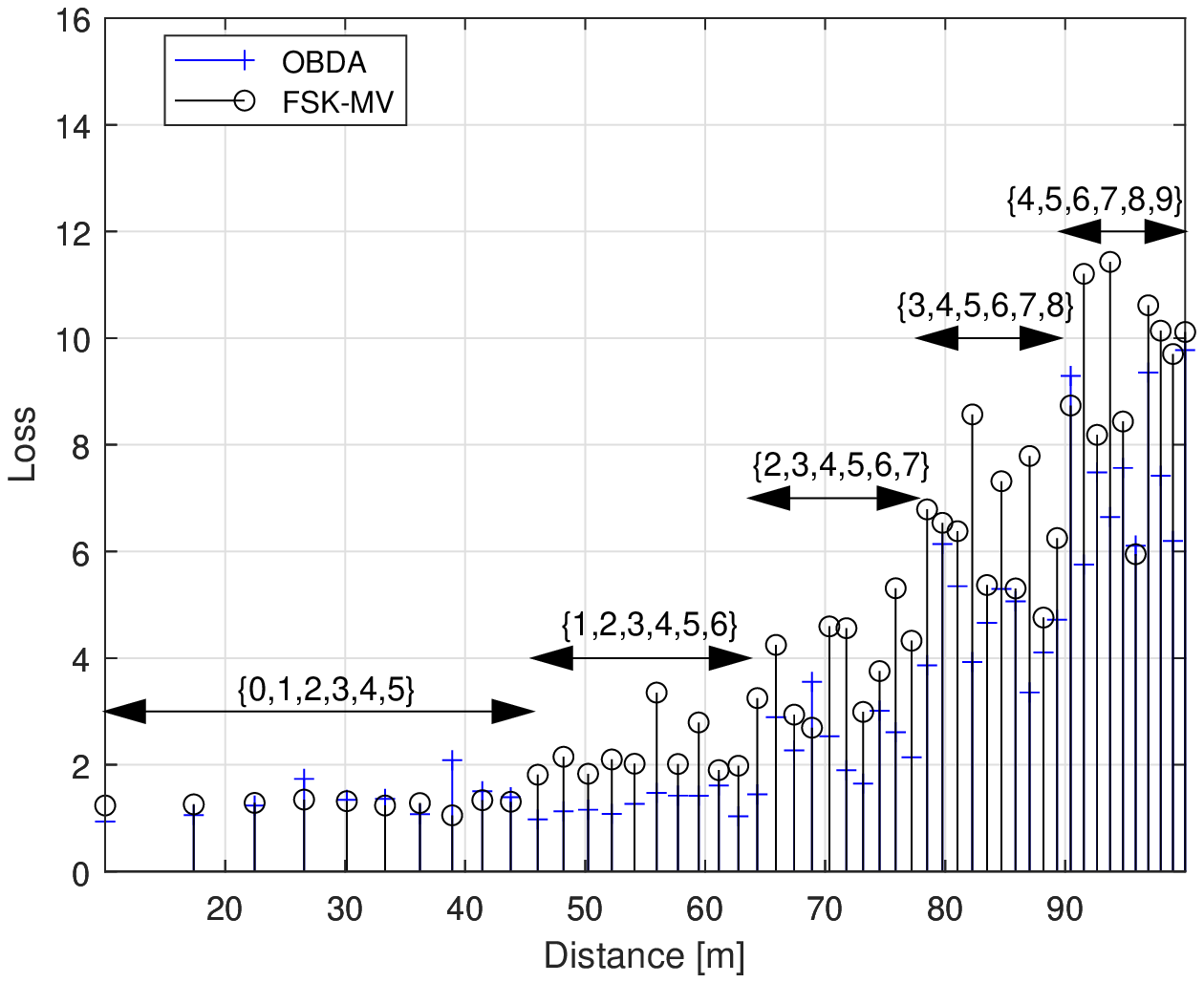}
		\label{subfig:ld_niid_aeff_two}}		
	\caption{Local loss versus link distance. For non-\ac{iid} data, the data samples are function of the locations of \acp{ED}. Since the received signal power of the cell-edge \acp{ED} are dominated by the nearby \acp{ED}, only data samples at the nearby \ac{ED} are learned. For this analysis, an ideal time synchronization is assumed in order to provide the results for \ac{OBDA}. The available labels are indicated as $\{\cdots\}$.}
	\label{fig:lossVsDistance}
\end{figure*}
For the numerical results, we consider the learning task of handwritten-digit recognition in a single cell with $\numberOfEdgeDevices=50$ \acp{ED} for $\minimumDistance=10$~meters and $\cellRadius=100$~meters. We assume that the path loss exponent is $\pathlossExponent=4$. To demonstrate the impact of the imperfect power control on distributed learning, we choose $\powerControl\in\{2,4\}$ and set the \ac{SNR}, i.e., $1/\noiseVariance$, to be $20$~dB at $\referenceDistance=10$~meters. 
The link distance between the $\indexED$th \ac{ED} and the \ac{ES} is set to $\distanceED[\indexED]=\sqrt{\minimumDistance^2 + (\indexED-1)(\cellRadius^2-\minimumDistance^2)/(\numberOfEdgeDevices-1)}$ based on \eqref{eq:linkDistance}.
For the fading channel, we consider ITU Extended Pedestrian A (EPA) with no mobility and regenerate the channels between the \ac{ES} and the \acp{ED} independently for each communication round to capture the long-term channel variations.
The subcarrier spacing is set to $15$~kHz. We use  $\numberOfActiveSubcarriers=1200$ subcarriers (i.e., the signal bandwidth is $18$~MHz). In the case of imperfect time synchronization, we assume that the difference between time of arriving \ac{ED} signals is maximum $\syncError=55.6$ ns and the  synchronization uncertainty at the \ac{ES} is $\Nerror=3$ samples. Otherwise, these parameters are set to $0$.

For the local data at the \acp{ED}, we use the MNIST database that contains labeled handwritten-digit images size of $28\times28$ from digit 0 to digit 9\footnote{For \ac{FEEL}, the data samples are generated at the \acp{ED}. We distribute the data samples in the MNIST database to the \acp{ED} to generate representative results for \ac{FEEL}.}. We consider both \ac{iid} data and non-\ac{iid} data in the cell. To prepare the data, we first choose $|\completeData|=25000$ training images from the database, where each digit has distinct $2500$ images.  
For the scenario with the \ac{iid} data, we assume that each \ac{ED} has $50$ distinct images for each digit. For the scenario with  the non-\ac{iid} data,   we assume that the distribution of the images depends on the locations of the \acp{ED} to test the \ac{FEEL} in a more challenging scenario. To this end, we divide the cell into 5 areas with concentric circles and the \acp{ED} located in $\indexArea$th area have the data samples with the labels $\{\indexArea-1,\indexArea,1+\indexArea,2+\indexArea,3+\indexArea,4+\indexArea\}$ for $\indexArea\in\{1,\mydots,5\}$.  Hence, the availability of the labels gradually changes based on the link distance. The areas between two adjacent concentric circles are identical and the number of \acp{ED} in each area is $10$. The  \ac{iid} and non-\ac{iid} data distributions are illustrated in \figurename~\ref{fig:niid}.

For the model, we consider a \ac{CNN}  that includes one $5\times5$ and two $3\times3$ convolutional layers, where each of them is followed by a  batch normalization layer and \ac{ReLU} activation follow each of them. All convolutional layers have $20$ filters. After the third \ac{ReLU}, a fully-connected layer with 10 units and a softmax layer are utilized. At the input layer, no normalization is applied. Our model, outline in Table~\ref{table:layout}, has $\numberOfModelParameters=123090$ learnable parameters, which corresponds to $\numberOfOFDMSymbols=206$ and $\numberOfOFDMSymbols=52$ OFDM symbols for the \ac{FSK-MV} and \ac{OBDA}  \cite{Guangxu_2021}, respectively.  For \ac{TCI}, the truncation threshold is $0.2$ and we assume that \ac{CSI} is available at the \acp{ED}. For the update rule, the learning rate is set to $0.01$. The batch size $\batchSize$ is set to $64$. For the test accuracy calculations, we use $10000$ test samples available in the MNIST database.

In \figurename~\ref{fig:testAcc}, we provide the test accuracy results for \ac{iid}/non-\ac{iid} data in the cell by taking time-synchronization errors and imperfect power control. For the same configurations, we provide the local loss values at the \acp{ED} as function of link distance in \figurename~\ref{fig:lossVsDistance} after $\communicationRounds=500$ communication rounds. 
In \figurename~\ref{fig:testAcc}\subref{subfig:acc_iid_aeff_zero} and \figurename~\ref{fig:testAcc}\subref{subfig:acc_iid_aeff_two}, we consider the \ac{iid} data in the cell. We evaluate the scenarios with the non-\ac{iid} data in \figurename~\ref{fig:testAcc}\subref{subfig:acc_niid_aeff_zero} and \figurename~\ref{fig:testAcc}\subref{subfig:acc_niid_aeff_two}. For \figurename~\ref{fig:testAcc}\subref{subfig:acc_iid_aeff_zero}, the power alignment at the \ac{ES} is assumed to be perfect (i.e., $\effectivePathLossExponent=0$). The results in this figure indicate that \ac{OBDA} works well when the time synchronization is ideal and the \ac{CSI} is available at the \acp{ED}. However, OBDA without \ac{TCI} or its utilization under imperfect time synchronization cause drastic reductions in the performance. On the other hand, the \ac{FSK-MV} is robust against the time-synchronization errors and result a high test accuracy without using \ac{CSI} at the \acp{ED} as it is based on non-coherent detection and dedicates two orthogonal resources to indicate the sign of the gradient.  In \figurename~\ref{fig:testAcc}\subref{subfig:ld_iid_aeff_two}-\subref{subfig:ld_niid_aeff_two}, we observe the same trends for \ac{OBDA} and \ac{FSK-MV}. However, the maximum test accuracy is highly affected by the data distribution and the power control. 
In \figurename~\ref{fig:testAcc}\subref{subfig:acc_iid_aeff_two}, the power alignment at the \ac{ES} is not ideal (i.e., $\effectivePathLossExponent=2$). Although the test accuracy with \ac{OBDA} with \ac{TCI} (with ideal synchronization) or \ac{FSK-MV} (with/without ideal synchronization) reaches to 95\%, \figurename~\ref{fig:lossVsDistance}\subref{subfig:ld_iid_aeff_two} indicates the local losses increase at the \acp{ED} as compared to the ones in \figurename~\ref{fig:lossVsDistance}\subref{subfig:ld_iid_aeff_zero}. In this scenario, the distributed learning exploits the \ac{iid}-data in the cell, which also benefits to the cell-edge \acp{ED} that have the similar data distributions to the ones at the nearby \acp{ED}. In \figurename~\ref{fig:testAcc}\subref{subfig:acc_niid_aeff_zero}, we see the impact of the non-\ac{iid} data on the test accuracy. Although the power alignment is ideal in this case, the maximum test accuracy reduces to $75\%$ from $95\%$. We observe more degradation in accuracy in \figurename~\ref{fig:testAcc}\subref{subfig:acc_niid_aeff_two}, where the power control is not ideal. In \figurename~\ref{fig:lossVsDistance}\subref{subfig:ld_niid_aeff_zero} and  \figurename~\ref{fig:lossVsDistance}\subref{subfig:ld_niid_aeff_two}, we can identify the digits that are not learned well. In the case of ideal power control, based on \figurename~\ref{fig:lossVsDistance}\subref{subfig:ld_niid_aeff_zero}, we observe that the digit 0 and the digit 9 are not learned well since these digits are available in less number of \acp{ED} as compared to other digits. Hence, the \ac{MV} is highly biased. A similar issue arises when the power control is not perfect. As shown in \figurename~\ref{fig:lossVsDistance}\subref{subfig:ld_niid_aeff_two}, the local loss function tend to increase with the distance, i.e., the cell-edge \acp{ED}'s data are not learned. As the cell-edge \acp{ED}' received signal powers are weak as compared the ones for the nearby \acp{ED}, the \ac{MV} is again biased toward the nearby \acp{ED} local data. Therefore, the digits available at the cell-edge \acp{ED}, e.g.,  digits 6, 7, 8, and 9, are not learned well. Both issues in the case of non-\ac{iid} data indicate that  an adaptive learning strategy that takes the bias in the \ac{MV} into account (e.g., through an adaptive \ac{ED} selection or a power control based on the label distribution) is needed for achieving a higher test accuracy.

\begin{figure}[t]
	\centering
	{\includegraphics[width =3.0in]{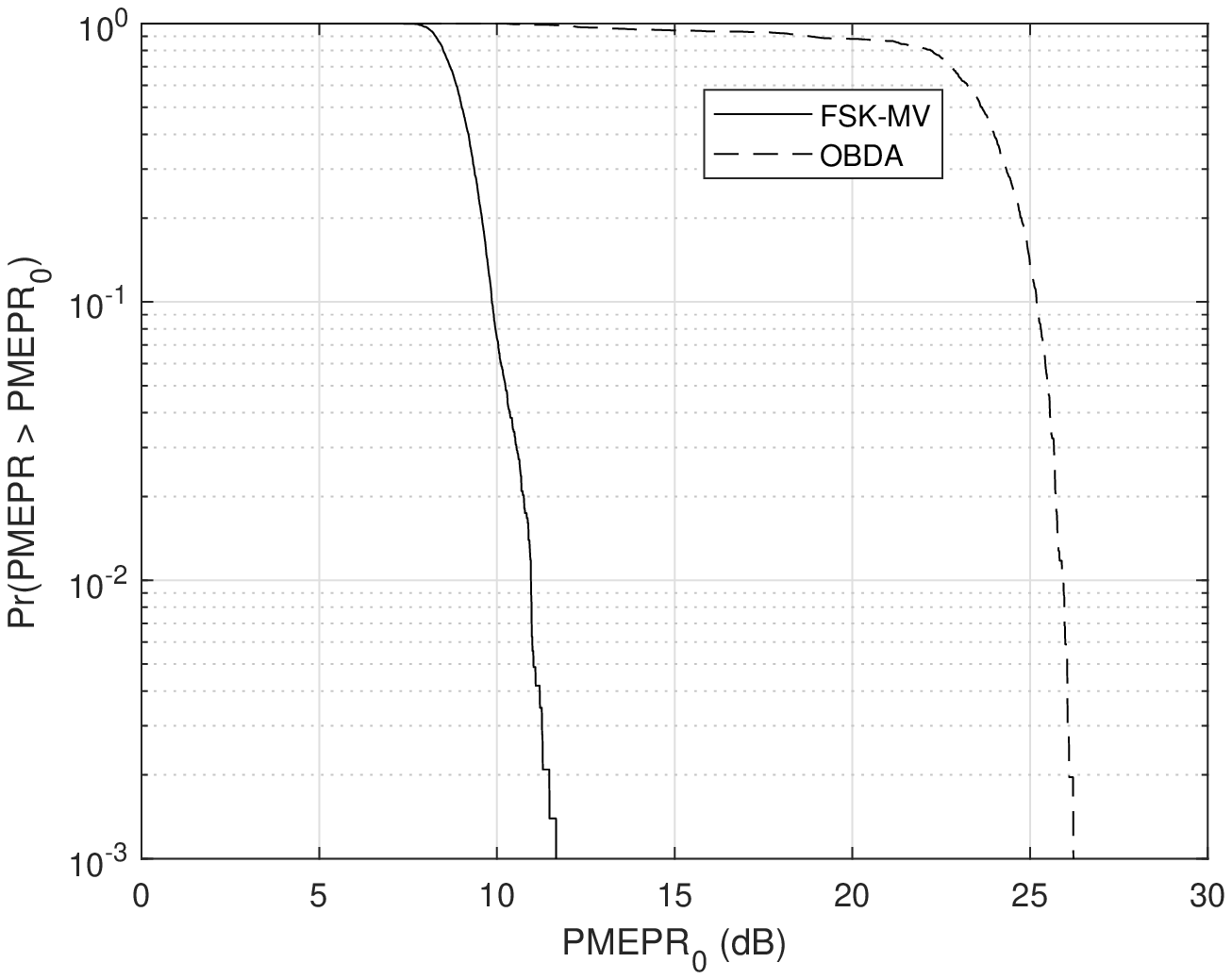}
	} 
	\caption{PMEPR distributions. The randomization symbols in \ac{FSK-MV} lowers \ac{PMEPR}. }
	\label{fig:pmepr}
\end{figure}
Finally, we compare the \ac{PMEPR} distributions in \figurename~\ref{fig:pmepr} for \ac{OBDA} and \ac{FSK-MV}. Since the proposed scheme introduces randomness in the frequency domain with the randomization symbols, it exhibits a similar behavior to a typical \ac{OFDM} transmission in terms of \ac{PMEPR}. On the other hand, the \ac{OBDA} can cause substantially high \ac{PMEPR} for \ac{OFDM} as the signs of the gradients can be highly-correlated.

\section{Concluding Remarks}
\label{sec:conclusion}
In this study, we propose an effective \ac{OAC} scheme for \ac{FEEL}. The proposed scheme relies on the distributed learning by the \ac{MV} with the \ac{signSGD} in fading channel. As compared to the state-of-the-art solutions on \ac{OAC}, it uses different subcarriers and/or OFDM symbols to indicate the sign of the local stochastic gradients. Thus, it allows the receiver at the \ac{ES} to detect the \ac{MV} with a non-coherent detector and eliminates the need for \ac{CSI} at the \acp{ED} by exploiting the non-coherent energy accumulation on the subcarriers.  We also prove the convergence of the distributed learning  by taking path loss, power control, and cell size into account. 
Through simulations, we demonstrate that the proposed method can provide a high test accuracy in fading channel even when the power control and the time synchronization are imperfect while resulting in an acceptable \ac{PMEPR} distribution at the expense of a larger number of time and frequency resources. We also provide insights into the scenarios where local data distribution depends on the locations of the \acp{ED} and demonstrate the impact of non-\ac{iid} data on the distributed learning when the power control is not ideal. Our results indicate that adaptive learning methods that consider the bias in the \ac{MV} due to the non-\ac{iid} data and/or imperfect power control are required for achieving a higher test accuracy.


\appendix[Proof of Theorem~\ref{th:convergence}]
\begin{IEEEproof}
	The proof of Theorem~\ref{th:convergence} relies on a well-known strategy of relating the norm of the gradient of the loss function $ \lossFunctionGlobal[\modelParameters]$ to the expected improvement made in a single step as described in \cite{Bernstein_2018}. Let $\globalGradient[\indexCommunicationRound]$ be the gradient of $\lossFunctionGlobal[{\modelParametersAtIteration[\indexCommunicationRound]}]$ (i.e., the true gradient                                                                                                                                                                                                                                                                                                                                                                                                                                                                                                                                                                                                                                                     ). By using Assumption~2 and using \eqref{eq:detector}, we can state that
	\begin{align}
		\lossFunctionGlobal[{\modelParametersAtIteration[\indexCommunicationRound+1]}]& - \lossFunctionGlobal[{\modelParametersAtIteration[\indexCommunicationRound]}]\le -\learningRate{\globalGradient[\indexCommunicationRound]}^{\rm T}\majorityVote[\indexCommunicationRound] + \frac{\learningRate^2}{2}\norm{\nonnegativeConstants}_1\nonumber\\
		=&-\learningRate\norm{\globalGradient[\indexCommunicationRound]}_1+\frac{\learningRate^2}{2}\norm{\nonnegativeConstants}_1\nonumber\\&+2\learningRate\sum_{\indexGradient=1}^{\numberOfModelParameters}|\globalGradientElement[\indexCommunicationRound][\indexGradient]| \indicatorFunction[{\signNormal[{\deltaVectorAtIteration[\indexCommunicationRound][\indexGradient]}]\neq \signNormal[{\globalGradientElement[\indexCommunicationRound][\indexGradient]}]}]\nonumber~.
	\end{align}	
	Therefore,
	\begin{align}
		&\expectationOperator[{	\lossFunctionGlobal[{\modelParametersAtIteration[\indexCommunicationRound+1]}] - \lossFunctionGlobal[{\modelParametersAtIteration[\indexCommunicationRound]}]|\modelParametersAtIteration[\indexCommunicationRound]}][] \le  -\learningRate\norm{\globalGradient[\indexCommunicationRound]}_1+\frac{\learningRate^2}{2}\norm{\nonnegativeConstants}_1\nonumber\\&~~~~~~~~~~~~~~+\underbrace{2\learningRate\sum_{\indexGradient=1}^{\numberOfModelParameters}|\globalGradientElement[\indexCommunicationRound][\indexGradient]| \underbrace{\probability[{\signNormal[{\deltaVectorAtIteration[\indexCommunicationRound][\indexGradient]}]\neq \signNormal[{\globalGradientElement[\indexCommunicationRound][\indexGradient]}]}]\nonumber}_{\triangleq\probabilityIncorrect[\indexGradient]}}_{\text{Stochasticity-induced error}}~.
	\end{align}
	The main challenge is to obtain an upper bound on the stochasticity-induced error. To address this, assume that $\signNormal[{\globalGradientElement[\indexCommunicationRound][\indexGradient]}]=1$. Let $\numberOfEDsWithCorrectChoice$ be a random variable for counting the number of EDs with the correct decision, i.e., $\signNormal[{\globalGradientElement[\indexCommunicationRound][\indexGradient]}]=1$. The random variable $\numberOfEDsWithCorrectChoice$ can  then be model as the sum of $\numberOfEdgeDevices$ independent Bernoulli trials, i.e., a binomial variable with the success and failure probabilities given by
	\begin{align}
		\correctDecision[\indexGradient]\triangleq\probability[{\signNormal[{\localGradientElement[\indexED,\indexGradient][\indexCommunicationRound]}]= \signNormal[{\globalGradientElement[\indexCommunicationRound][\indexGradient]}]}]\nonumber~,\\
		\incorrectDecision[\indexGradient]\triangleq\probability[{\signNormal[{\localGradientElement[\indexED,\indexGradient][\indexCommunicationRound]}]\neq \signNormal[{\globalGradientElement[\indexCommunicationRound][\indexGradient]}]}]\nonumber~,
	\end{align}
	respectively, for all $\indexED$. This implies that
	\begin{align}
		\probabilityIncorrect[\indexGradient]=
		\sum_{\numberOFEDsForOptionOne=0}^{\numberOfEdgeDevices}
		\probability[{\signNormal[{\deltaVectorAtIteration[\indexCommunicationRound][\indexGradient]}]\neq 1}|\numberOfEDsWithCorrectChoice=\numberOFEDsForOptionOne]\probability[\numberOfEDsWithCorrectChoice=\numberOFEDsForOptionOne]~, \nonumber
	\end{align}
	where
	$\probability[\numberOfEDsWithCorrectChoice=\numberOFEDsForOptionOne] = \binom{\numberOfEdgeDevices}{\numberOFEDsForOptionOne}\correctDecision[\indexGradient]^{\numberOFEDsForOptionOne}\incorrectDecision[\indexGradient]^{\numberOfEdgeDevices-\numberOFEDsForOptionOne}$.
	To calculate $\probability[{\signNormal[{\deltaVectorAtIteration[\indexCommunicationRound][\indexGradient]}]\neq 1}|\numberOfEDsWithCorrectChoice=\numberOFEDsForOptionOne]$, we use the distribution of $\deltaVectorAtIteration[\indexCommunicationRound][\indexGradient]$, which can be obtained by using the properties of exponential random variables as
	\begin{align}
		f(\deltaVectorAtIteration[\indexCommunicationRound][\indexGradient]) = \begin{cases} 
			\frac{\constante^{-\frac{\deltaVectorAtIteration[\indexCommunicationRound][\indexGradient]}{\meanOptionTwo}}}{\meanOptionOne+\meanOptionTwo}, & \deltaVectorAtIteration[\indexCommunicationRound][\indexGradient]\le 0 \\
			\frac{\constante^{-\frac{\deltaVectorAtIteration[\indexCommunicationRound][\indexGradient]}{\meanOptionOne}}}{\meanOptionOne+\meanOptionTwo}, & \deltaVectorAtIteration[\indexCommunicationRound][\indexGradient]>0 
		\end{cases}~.
		\label{eq:lappro}
	\end{align}
	Thus, by integrating \eqref{eq:lappro} with respect to $\deltaVectorAtIteration[\indexCommunicationRound][\indexGradient]$,
	\begin{align}
		\probability[{\signNormal[{\deltaVectorAtIteration[\indexCommunicationRound][\indexGradient]}]\neq 1}|\numberOfEDsWithCorrectChoice=\numberOFEDsForOptionOne]& = \frac{\meanOptionTwo}{\meanOptionOne+\meanOptionTwo}\nonumber
		\\&=\frac{(\numberOfEdgeDevices-\numberOFEDsForOptionOne)+1/\effectiveSNR}{\numberOfEdgeDevices+2/\effectiveSNR}.
		\label{eq:probLapResult}
	\end{align}
	Hence, by using \eqref{eq:probLapResult} and the properties of binomial coefficients
	\begin{align}
		\probabilityIncorrect[\indexGradient]&=\sum_{\numberOFEDsForOptionOne=0}^{\numberOfEdgeDevices}
		\frac{(\numberOfEdgeDevices-\numberOFEDsForOptionOne)+1/\effectiveSNR}{1+2/\effectiveSNR}\binom{\numberOfEdgeDevices}{\numberOFEDsForOptionOne}\correctDecision[\indexGradient]^{\numberOFEDsForOptionOne}\incorrectDecision[\indexGradient]^{\numberOfEdgeDevices-\numberOFEDsForOptionOne}\nonumber\\
		&=\frac{\frac{1}{\effectiveSNR\numberOfEdgeDevices}}{1+\frac{2}{\numberOfEdgeDevices\effectiveSNR}}+\frac{\incorrectDecision[\indexGradient]}{1+\frac{2}{\effectiveSNR\numberOfEdgeDevices}}~.
	\end{align}
	Under Assumption 2 and Assumption 3, by using the derivations in \cite{Bernstein_2018}, it can be shown that $\incorrectDecision[\indexGradient]\le\frac{\sqrt{2}\varianceBoundEle[\indexGradient]}{3|\globalGradientElement[\indexCommunicationRound][\indexGradient]|\sqrt{\batchSize}}$. Hence, an upper bound on the stochasticity-induced error can be obtained as
	\begin{align}
		\sum_{\indexGradient=1}^{\numberOfModelParameters}|\globalGradientElement[\indexCommunicationRound][\indexGradient]|\probabilityIncorrect[\indexGradient]\le  \frac{\frac{1}{\effectiveSNR\numberOfEdgeDevices}}{1+\frac{2}{\numberOfEdgeDevices\effectiveSNR}}\norm{\globalGradient[\indexCommunicationRound]}_1 +\frac{1}{\sqrt{\batchSize}} \frac{\sqrt{2}/3}{1+\frac{2}{\numberOfEdgeDevices\effectiveSNR}}\norm{\varianceBound}_1~.\nonumber
	\end{align}
	Based on Assumption~1, 
	\begin{align}
		&\lossFunctionGlobal[{\modelParametersAtIteration[0]}]-\lossFunctionGlobalMinimum\ge \lossFunctionGlobal[{\modelParametersAtIteration[0]}]-\expectationOperator[{\lossFunctionGlobal[{\modelParametersAtIteration[\communicationRounds]}]}][]\nonumber\\&=\expectationOperator[{\sum_{\indexCommunicationRound=0}^{\communicationRounds-1}\lossFunctionGlobal[{\modelParametersAtIteration[\indexCommunicationRound]}] - \lossFunctionGlobal[{\modelParametersAtIteration[\indexCommunicationRound+1]}]}][]\nonumber\\
		&\ge
		\expectationOperator[{	\sum_{\indexCommunicationRound=0}^{\communicationRounds-1}\frac{\learningRate}{1+\frac{2}{\numberOfEdgeDevices\effectiveSNR}}\norm{\globalGradient[\indexCommunicationRound]}_1-\frac{\learningRate^2}{2}\norm{\nonnegativeConstants}_1 -\frac{\learningRate}{\sqrt{\batchSize}} \frac{2\sqrt{2}/3}{1+\frac{2}{\numberOfEdgeDevices\effectiveSNR}}\norm{\varianceBound}_1  }][]~.
		\label{eq:finaleq}
	\end{align}
	By rearranging the terms in \eqref{eq:finaleq} and using the expressions for $\batchSize$ and $\learningRate$, \eqref{eq:convergence} is reached.
\end{IEEEproof}

\acresetall

\bibliographystyle{IEEEtran}
\bibliography{references}

\begin{thebibliography}{10}
\providecommand{\url}[1]{#1}
\csname url@samestyle\endcsname
\providecommand{\newblock}{\relax}
\providecommand{\bibinfo}[2]{#2}
\providecommand{\BIBentrySTDinterwordspacing}{\spaceskip=0pt\relax}
\providecommand{\BIBentryALTinterwordstretchfactor}{4}
\providecommand{\BIBentryALTinterwordspacing}{\spaceskip=\fontdimen2\font plus
\BIBentryALTinterwordstretchfactor\fontdimen3\font minus
  \fontdimen4\font\relax}
\providecommand{\BIBforeignlanguage}[2]{{%
\expandafter\ifx\csname l@#1\endcsname\relax
\typeout{** WARNING: IEEEtran.bst: No hyphenation pattern has been}%
\typeout{** loaded for the language `#1'. Using the pattern for}%
\typeout{** the default language instead.}%
\else
\language=\csname l@#1\endcsname
\fi
#2}}
\providecommand{\BIBdecl}{\relax}
\BIBdecl

\bibitem{gafni2021federated}
\BIBentryALTinterwordspacing
T.~Gafni, N.~Shlezinger, K.~Cohen, Y.~C. Eldar, and H.~V. Poor, ``Federated
  learning: A signal processing perspective,'' 2021. [Online]. Available:
  \url{arXiv:2103.17150}
\BIBentrySTDinterwordspacing

\bibitem{chen2021distributed}
M.~Chen, D.~Gündüz, K.~Huang, W.~Saad, M.~Bennis, A.~V. Feljan, and
  H.~Vincent~Poor, ``Distributed learning in wireless networks: Recent progress
  and future challenges,'' \emph{IEEE J. Sel. Areas Commun.}, pp. 1--26, 2021.

\bibitem{hellstrom2020wireless}
H.~Hellstrom, J.~M.~B. da~Silva~Jr, V.~Fodor, and C.~Fischione, ``Wireless for
  machine learning,'' 2020.

\bibitem{Goldenbaum_2013}
M.~Goldenbaum, H.~Boche, and S.~Stańczak, ``Harnessing interference for analog
  function computation in wireless sensor networks,'' \emph{IEEE Trans. Signal
  Process.}, vol.~61, no.~20, pp. 4893--4906, Oct. 2013.

\bibitem{Wanchun_2020}
W.~Liu, X.~Zang, Y.~Li, and B.~Vucetic, ``Over-the-air computation systems:
  Optimization, analysis and scaling laws,'' \emph{IEEE Trans. Wireless
  Commun.}, vol.~19, no.~8, pp. 5488--5502, Aug. 2020.

\bibitem{Nazer_2007}
B.~Nazer and M.~Gastpar, ``Computation over multiple-access channels,''
  \emph{IEEE Trans. Inf. Theory}, vol.~53, no.~10, pp. 3498--3516, Oct. 2007.

\bibitem{Guangxu_2020}
G.~Zhu, Y.~Wang, and K.~Huang, ``Broadband analog aggregation for low-latency
  federated edge learning,'' \emph{IEEE Trans. Wireless Commun.}, vol.~19,
  no.~1, pp. 491--506, Jan. 2020.

\bibitem{sery2020overtheair}
\BIBentryALTinterwordspacing
T.~Sery, N.~Shlezinger, K.~Cohen, and Y.~C. Eldar, ``Over-the-air federated
  learning from heterogeneous data,'' 2020. [Online]. Available:
  \url{arXiv:2009.12787}
\BIBentrySTDinterwordspacing

\bibitem{Amiri_2020}
M.~M. Amiri and D.~Gündüz, ``Federated learning over wireless fading
  channels,'' \emph{IEEE Trans. Wireless Commun.}, vol.~19, no.~5, pp.
  3546--3557, Feb. 2020.

\bibitem{Guangxu_2021}
G.~Zhu, Y.~Du, D.~Gündüz, and K.~Huang, ``One-bit over-the-air aggregation
  for communication-efficient federated edge learning: Design and convergence
  analysis,'' \emph{IEEE Trans. Wireless Commun.}, vol.~20, no.~3, pp.
  2120--2135, Nov. 2021.

\bibitem{Bernstein_2018}
J.~Bernstein, Y.-X. Wang, K.~Azizzadenesheli, and A.~Anandkumar, ``sign{SGD}:
  Compressed optimisation for non-convex problems,'' in \emph{Proc. in
  International Conference on Machine Learning}, vol.~80.\hskip 1em plus 0.5em
  minus 0.4em\relax Proceedings of Machine Learning Research, 10--15 Jul 2018,
  pp. 560--569.

\bibitem{Yang_2020}
K.~Yang, T.~Jiang, Y.~Shi, and Z.~Ding, ``Federated learning via over-the-air
  computation,'' \emph{IEEE Trans. Wireless Commun.}, vol.~19, no.~3, pp.
  2022--2035, 2020.

\bibitem{Amiria_2021}
M.~M. Amiria, T.~M. Duman, D.~Gündüz, S.~R. Kulkarni, and H.~Vincent~Poor,
  ``Collaborative machine learning at the wireless edge with blind
  transmitters,'' \emph{IEEE Trans. Wireless Commun.}, pp. 1--1, Mar 2021.

\bibitem{10.5555/3294673}
E.~Dahlman, S.~Parkvall, and J.~Skold, \emph{{5G NR}: The Next Generation
  Wireless Access Technology}, 1st~ed.\hskip 1em plus 0.5em minus 0.4em\relax
  USA: Academic Press, Inc., 2018.

\bibitem{sahinCCNC_2022_submit}
A.~{\c{S}ahin}, B.~{Everette}, and S.~{Hoque}, ``Over-the-air computation with
  {DFT-spread OFDM} for federated edge learning,'' in \emph{Proc. IEEE Wireless
  Communications and Networking Conference (WCNC) (submitted)}, Apr. 2022, pp.
  1--6.

\bibitem{Zeng_2020}
T.~Zeng, O.~Semiari, M.~Mozaffari, M.~Chen, W.~Saad, and M.~Bennis, ``Federated
  learning in the sky: Joint power allocation and scheduling with {UAV}
  swarms,'' in \emph{Proc. IEEE International Conference on Communications
  (ICC)}, 2020, pp. 1--6.

\bibitem{Jawhar_2019}
Y.~A. Jawhar, L.~Audah, M.~A. Taher, K.~N. Ramli, N.~S.~M. Shah, M.~Musa, and
  M.~S. Ahmed, ``A review of partial transmit sequence for {PAPR} reduction in
  the {OFDM} systems,'' \emph{IEEE Access}, vol.~7, pp. 18\,021--18\,041, 2019.

\end{thebibliography}

\end{document}